\documentclass{pasa}%

\usepackage{graphicx}

\usepackage{gensymb}
\usepackage{amssymb}
\usepackage{threeparttable}
\usepackage{upgreek}

\title[Snapshot astrometric accuracy]{Astrometric accuracy of snapshot Fast Radio Burst localisations with ASKAP}

\author[Cherie K. Day et al.]
{Cherie K. Day$^{1,2}$\thanks{contact: cday@swin.edu.au}, 
Adam~T.~Deller$^1$, 
Clancy~W.~James$^3$,
Emil Lenc$^2$, 
Shivani Bhandari$^2$, 
R.~M.~Shannon$^1$, 
and
Keith~W.~Bannister$^2$
\affil{$^{1}$Centre for Astrophysics and Supercomputing, Swinburne University of Technology, Hawthorn VIC 3122, Australia}
\affil{$^{2}$CSIRO, Space and Astronomy, PO Box 76, Epping NSW 1710 Australia}
\affil{$^{3}$International Centre for Radio Astronomy Research, Curtin University, Bentley, WA 6102, Australia}
}%

\jid{PASA}
\doi{10.1017/pas.\the\year.xxx}
\jyear{\the\year}

\usepackage{aas_macros}
\usepackage{hyperref} 
\hypersetup{colorlinks,citecolor=blue,linkcolor=blue,urlcolor=blue}

\begin{document}

\begin{frontmatter}
\maketitle

\begin{abstract}
The recent increase in well-localised fast radio bursts (FRBs) has facilitated in-depth studies of global FRB host properties, the source circumburst medium, and the potential impacts of these environments on the burst properties. The Australian Square Kilometre Array Pathfinder (ASKAP) has localised 11 FRBs with sub-arcsecond to arcsecond precision, leading to sub-galaxy localisation regions in some cases and those covering much of the host galaxy in others. The method used to astrometrically register the FRB image frame for ASKAP, in order to align it with images taken at other wavelengths, is currently limited by the brightness of continuum sources detected in the short-duration (`snapshot') voltage data captured by the Commensal Real-Time ASKAP Fast Transients (CRAFT) software correlator, which are used to correct for any frame offsets due to imperfect calibration solutions and estimate the accuracy of any required correction. In this paper, we use dedicated observations of bright, compact radio sources in the low- and mid-frequency bands observable by ASKAP to investigate the typical astrometric accuracy of the positions obtained using this so-called `snapshot' technique. Having captured these data with both the CRAFT software and ASKAP hardware correlators, we also compare the offset distributions obtained from both data products to estimate a typical offset between the image frames resulting from the differing processing paths, laying the groundwork for future use of the longer-duration, higher signal-to-noise ratio (S/N) data recorded by the hardware correlator. We find typical offsets between the two frames of $\sim 0.6$ and $\sim 0.3$ arcsec in the low- and mid-band data, respectively, for both RA and Dec. We also find reasonable agreement between our offset distributions and those of the published FRBs. We detect only a weak dependence in positional offset on the relative separation in time and elevation between target and calibrator scans, with the trends being more pronounced in the low-band data and in Dec. Conversely, the offsets show a clear dependence on frequency in the low band, which we compare to the frequency-dependent Dec.\ offsets found in FRB~200430. In addition, we present a refined methodology for estimating the overall astrometric accuracy of CRAFT FRBs.

\end{abstract}

\end{frontmatter}

\section{INTRODUCTION}
\label{sec:intro}

Fast radio bursts (FRBs) are highly energetic, of order $\upmu$s to ms duration bursts of emission arising out to cosmological distances. While several hundred FRBs have been detected to date\footnote{See, for example, the Transient Name Server at https://www.wis-tns.org/; see also \cite{CHIME9repeaters2020ApJ...891L...6F} for a mention of the upcoming Canadian Hydrogen Intensity Mapping Experiment (CHIME) catalogue of FRBs}, their emission mechanism(s) and progenitor(s) are as yet unknown. Precise localisation of FRB sources is a critical step toward discriminating between viable pathways of FRB creation. Such localisations by the Karl G. Jansky Very Large Array (VLA), the Australian Square Kilometre Array Pathfinder (ASKAP), the Deep Synoptic Array (DSA-10), and the European VLBI Network (EVN) \citep[see, e.g.,][]{Chatterjee17,Bannister565,Ravi_2019,Marcote2020Natur.577..190M,Law2020ApJ...899..161L} have facilitated not only host galaxy identification, which requires localisations of $\lesssim$ a few arcsec precision (a requirement that gets increasingly stringent at higher redshift), but also in-depth studies relating burst properties to local environments \citep[e.g.,][]{Michilli2018Natur.553..182M,Tendulkar2021ApJ...908L..12T}, offering clues about the nature of the emission mechanism and progenitor.

The sub-arcsecond to arcsecond localisation of 14 FRBs \citep[see, e.g., ][]{Prochaska231, Macquart2020_DMz,Marcote2020Natur.577..190M} has yielded in-depth studies of the global host galaxy properties \citep{Bhandari2020_hostgalaxies,Heintz_hostII_2020ApJ...903..152H}. Investigating the varied host properties and offset distributions of FRBs, they determine which of the proposed progenitors are common to all host galaxy types, thereby constraining the likelihood of several proposed common sources of FRBs (i.e., when taking both repeating and apparently non-repeating FRBs to be from a single population). They reject active galactic nuclei based on the galactic centre offsets of several FRBs \citep{Bhandari2020_hostgalaxies} and find that galaxies typically hosting short gamma-ray bursts (SGRBs) and core-collapse and Type Ia supernovae were favoured as common hosts over those hosting long gamma-ray bursts \citep[LGRBs,][]{Heintz_hostII_2020ApJ...903..152H}. While the studied sample thus far has shed some light on the origins of FRBs, a growing sample of both highly accurate and precise (i.e., sub-galaxy) positions will improve our understanding of the local environments of FRBs, further constraining the progenitor and emission mechanism models. Additionally, since increased localisation precision will help to constrain the contributions to the dispersion measure and rotation measure from both the host and circumburst media, this information can then be used to improve models of extragalactic contributions that are employed when using FRBs as probes of, for example, large scale structure or cosmology \citep[for discussions of potential uses and early results of FRBs as probes, see, e.g.][]{Walters_2018,Prochaska231,Macquart2020_DMz}.

For a galaxy with a redshift of $\rm 0.04 < z < 0.5$, a precision of 1 arcsec corresponds to a projected angular scale range\footnote{http://www.astro.ucla.edu/$\sim$wright/CosmoCalc.html \citep{Wright2006PASP..118.1711W}, where we have assumed standard Planck cosmology \citep{Plank2020A&A...641A...6P} -- i.e., $\mathrm{H_0 = 67.5}$ km s$^{-1}$ Mpc$^{-1}$, $\mathrm{\Omega_M = 0.315}$, and $\mathrm{\Omega_{\Lambda} = 0.685}$ -- and varied the redshift.} of $\sim 1 - 6$~kpc. \cite{Mannings2020} investigated the host galaxies of a sample of eight localised FRBs within this redshift range and found that the galaxies had a similar range of angular sizes. Given an image signal-to-noise (S/N) $\geq$50, the native localisation precision of ASKAP is $\sim0.2$~arcsec or better, but the systematic astrometric offsets related to imperfect calibration can degrade this by roughly an order of magnitude. Therefore, some ASKAP FRB localisations have yielded sub-galaxy positional information, while others have only been able to differentiate between potential hosts. Moreover, for these studies to be meaningful, the reliability of the estimated positional uncertainties is crucial since underestimation can lead to erroneous conclusions while overestimation results in losing the ability to infer local characteristics from the position.

In this work, we characterise the typical astrometric accuracy attainable with the snapshot localisation technique employed for ASKAP FRBs when observing in the Commensal Real-Time ASKAP Fast Transients \citep[CRAFT,][]{CRAFT_2010,Bannister565} survey mode (i.e., using the CRAFT software correlator data) and lay the groundwork for long-term improvements. We use dedicated ASKAP observations to characterise our \textit{a priori} calibration accuracy, simultaneously recording data with the ASKAP hardware correlator and the CRAFT software correlator. Throughout this paper, for simplicity, we will use ASKAP and CRAFT to denote data products or results specific to the primary ASKAP hardware correlator signal path \citep{Hotan2021PASA...38....9H} and the CRAFT software correlator signal path outputs, respectively, where we note that the common signal is split after beamforming. For simplicity, we use the designation `ASKAP hardware correlator' to refer to all of the signal chain steps that follow the beamformer through to the final cross-multiplication and accumulation \citep[as described in][]{Hotan2021PASA...38....9H}. We also compare these data to determine the typical systematic offset between the resultant ASKAP and CRAFT image frames. In Section \ref{sec:methods}, we describe the observations and the methods used to analyse the software and hardware correlator datasets. In Section \ref{sec:res_analysis}, we present a comparison of the positional offsets versus time, elevation, and angular separation from the calibrator scans to determine any dependence on these observational parameters. Section \ref{sec:res_analysis} also outlines our findings for the future use of continuum images formed from the ASKAP hardware correlator data to astrometrically register the FRB image frame. Finally, in Section \ref{sec:ast_vs_FRBdist}, we compare our results to the published CRAFT FRB offset distributions, investigate improvements to the current model used to estimate the total offsets and uncertainties applied to the FRB to tie its image frame to the third International Celestial Reference Frame \citep[ICRF3,][]{ICRF3_2018AGUFM.G42A..01G}, and discuss the frequency-dependent offsets found in FRB~200430.

\section{Methods} \label{sec:methods}

Astrometric positional accuracy and precision is significantly affected by the quality of the calibration solutions, and there are several factors that influence the accuracy of these solutions. Typically, radio observations alternate between the target and a nearby calibrator (i.e., a strong compact source of known position), and these calibrator data are used to model all contributions to the observation (e.g., all components of the total delay at each station). The \textit{a priori} calibration solutions derived are then interpolated (spatially and temporally) and applied to the target. FRB observations (and in particular those conducted commensally) are generally limited to observing a calibrator sometime after the detection is made and an observation can be scheduled. This can result in observations of calibrators that are significantly temporally and/or spatially separated from the target, which correspondingly impacts the accuracy of these solutions when interpolated and applied to the target. Moreover, any deviations between the model and the observations (e.g., those caused by station clock differences, the ionosphere, the troposphere, or changes in the propagation path) will lead to shifts in the phase of the visibilities, which act to shift and smear the source image, leading to a reduction in the measured source S/N and a systematic error in the recovered source position \citep[see, e.g., ][]{WhiteBooktaylor1999synthesis}.

As a result, three datasets are needed to perform FRB localisations with CRAFT data, as described in \cite{Bannister565}. Each of the following datasets is formed by correlating the captured voltage data using the Distributed FX (DiFX) software correlator \citep{Deller11}.
\begin{itemize}
    \item The so-called `gated' data are used to determine an initial FRB position. They are formed from an optimal slice of the full 3.1-second raw voltage data containing only the FRB. This maximises the S/N of the FRB image made from these data and hence achieves the lowest statistical uncertainty on the fitted FRB position.
    \item The `field' data are used to estimate and correct for the systematic error described above. These correlated data span the full 3.1-second duration of the voltage data in which the FRB was detected.
    \item The `calibrator' data are used to derive the phase and bandpass calibration solutions applied to both target datasets. These are generated from a separate voltage download which is made while pointing at a known calibrator source after the FRB has been detected.
\end{itemize}
Once the target datasets are calibrated,  frequency-averaged images are made for each using the Common Astronomy Software Applications \citep[\textsc{casa,}][]{CASA2007ASPC..376..127M} task \textsc{tclean}. The short duration of the captured target data likewise yields short-duration (so-called `snapshot') sampling of the ($u,v$)-plane, and so we refer to the resultant images as `snapshot images'. The position of any detected source in these images is fit with a 2-D Gaussian using the Astronomical Image Processing System \cite[\textsc{aips},][]{Greisen03} task \textsc{jmfit}, which estimates both the source position and the statistical uncertainty in the fit.

The positions of any background continuum sources in the field image are then compared to their counterpart positions, where the latter are obtained from data with calibration solutions that require minimal interpolation (i.e., from either a catalogue or dedicated observations at a similar spatial resolution). The source offsets found in this comparison can then be used to estimate and correct for the overall systematic shift in the image frame relative to the ICRF3, thereby registering the CRAFT image frame to that of the ICRF3. Assuming any systematic offsets present in the data due to imperfect calibration solutions manifest as merely translations of the image frame (i.e., taking direction-independent effects as dominant and, hence, explicitly neglecting any more complicated direction-dependent distortions, such as rotation or stretching), the simple weighted mean of the individual offsets is used for this final systematic offset correction. The corresponding uncertainty is then nominally estimated by taking the weighted mean of the quadrature-summed CRAFT and comparison source positional uncertainties \citep{Macquart2020_DMz}. For cases in which the scatter in the offsets of individual sources about the mean was clearly greater than expected based on the formal uncertainty in the positions of the individual sources, however, the scatter itself has been used to estimate the uncertainty in the systematic offset \citep[FRB~200430;][]{Heintz_hostII_2020ApJ...903..152H}.

The systematic uncertainty is typically dominated by the statistical uncertainty in the CRAFT-derived field source positions, noting the latter is directly dependent on the S/N of the detections. Thus, the ability to estimate the systematic error and the uncertainty in this estimation are limited by the S/N of the background sources. In order to maximise the astrometric accuracy of observations, it is therefore desirable to reduce the systematic errors caused by unmodelled delays to below the statistical, S/N-limited uncertainties of the measurements. In addition, reducing the latter will improve the overall precision of the final positions. \cite{Day2020} noted that the median statistical positional uncertainties in their sample of CRAFT FRBs are $\sim$(0.1,0.2)~arcsec for (RA,Dec.), and they argued that transferring the higher S/N calibration solutions derived from commensally captured ASKAP hardware correlator data would reduce the systematic uncertainties to roughly this precision.

While it is optimal to reduce the systematic errors via these refined calibration solutions, residual offsets between both the CRAFT and ASKAP frames and the ASKAP and reference frames will always exist. In the case of the former, while the CRAFT and ASKAP datasets are generated from identical data, differences in, e.g., the geometric models used by the two correlators, the approximations made in the associated signal processing steps, or in the calibration approaches used could lead to small offsets. For the latter, considering the ASKAP hardware correlator output compared to the ICRF3, \cite{RACS_I_2020PASA...37...48M} showed that an astrometric offset of up to $\sim1$~arcsec can exist, the source of which is not yet understood and is currently under investigation. In order to apply calibration solutions derived from the ASKAP hardware correlator data to the CRAFT data, each of these residual frame offsets must therefore be quantified, which can be accomplished in two ways. They can be determined in each case individually by comparing field images made from both CRAFT and ASKAP data to estimate the former and comparing the higher S/N ASKAP field source positions to a set of reference source positions to estimate the latter. Conversely, a global estimate of the typical residual offset between the CRAFT and ASKAP frames can be determined by observing several sources all over the sky, recording voltages with both correlators, and performing a comparison. Then, the ASKAP to reference frame offset can be estimated as per the current method, with the ASKAP field image replacing the CRAFT field image. We note that the ASKAP frame registration is expected to improve in the future, so once it can be shown to be well-registered, this latter step can be omitted since the ASKAP-CRAFT residual will likely significantly dominate the systematic offset and uncertainty.

\subsection{Observations} \label{sec:obs}

In order to determine the typical astrometric accuracy of the current snapshot method and evaluate the potential use of the ASKAP-derived calibration solutions, we must sample the effects of spatial and temporal deviations between the target and calibrator observations when interpolating the calibration solutions for use on the target data in both the CRAFT and ASKAP correlator data cases. To accomplish this, we observed a set of strong compact sources, with consistently high S/Ns, varying their spatial and temporal separations. We selected four sources within the ASKAP declination range \citep[$-90\degree$ to $+41\degree$, see][]{RACS_I_2020PASA...37...48M} from the VLA calibrator list with an ICRF3 (i.e., Very Long Baseline Array [VLBI] catalogue) counterpart: ICRF J155751.4$-$000150 (J1557), ICRF J191109.6$-$200655 (J1911), PKS 1934$-$638 (J1939), and PKS 2211$-$388 (J2214). These sources are specified as `P' class (i.e., strong compact sources) on the VLA in the A, B, C, and D configurations in both the L (1-2 GHz) and C (4-8 GHz) bands. This ensures that they are compact on (sub-)arcsecond scales, and hence the centroid measured by our ASKAP observations will be compatible with the catalogue position to high precision.

Two sets of observations were taken to characterise the typical astrometric accuracy in the low- and mid-frequency bands in which FRBs have been detected using the CRAFT data. Taken on two separate days, the low- and mid-band observations were, respectively, conducted using the CRAFT software correlator at central frequencies of 863.5~MHz and 1271.5~MHz, each with a total bandwidth of 336~MHz. The sources were observed at the selected central frequency in a repeated loop, being added in as they rose above ASKAP's horizon limit, across a range of elevation separations ($\lesssim70\degree$) and over a period of $\sim 8.2$~hours (mid-frequency band) and $\sim 5.6$~hours (low-frequency band).

Given the Phased Array Feed (PAF) used on the ASKAP dishes \citep{Hotan2021PASA...38....9H}, the performance of each of the 36 beams formed by the PAF is affected by its location within the PAF footprint. Therefore, the four sources were observed using beams located close to antenna boresight (beam 15), along the outer edge of the footprint (beam 30), and in between these two (beam 28) to determine any positional offset dependence on beam location. Each observation used the ASKAP `closepack36' configuration \cite[i.e., beams arranged in a hexagonal close pack configuration; see Figure 20 in][]{Hotan2021PASA...38....9H}, with a 45-degree PAF rotation to align the beams such that they tiled the sky in RA and Dec.\ The mid- and low-frequency bands respectively used a pitch (i.e., beam spacing) of 0.9~deg and 1.05~deg; we note that, at the lower frequencies, the beams can be spaced further apart and still retain reasonable sensitivity due to the larger beam size.

On UTC 2020 October 23, the four sources were observed in the mid band using a sub-array with a total of 17 antennas. Due to issues with voltage downloads for data from one antenna, this antenna was replaced during the observing run with another to retain the same array size. However, in order to maintain a consistent set of antennas throughout the observation, the two partially used antennas were subsequently removed during processing, reducing the array to 16 antennas. This sub-array has baselines ranging from 80~m to 5038~m. After removing any failed scans, the final data set contained 27 scans on beam 15, 28 scans on beam 28, and 27 scans on beam 30.

On UTC 2021 January 13, three of the four sources (J1557, J1939, and J2214) were observed in the low band using a sub-array of 17 antennas, with baselines ranging from 27~m to 5931~m. Since the source J1911 was within $10\degree$ of the Sun (i.e., within the Sun avoidance limits set for ASKAP), it could not be included in this observation. Upon processing the data, three J1939 scans (one per beam) were discovered to have voltage dropouts (i.e., where a download has failed and no data exist) in various blocks of the observed band. These scans were therefore removed from the processing. With this and the removal of all other scans with any voltage download issues, the final data set contained 18 scans on beam 15, 19 scans on beam 28, and 19 scans on beam 30.

The target sources were simultaneously recorded with the ASKAP hardware correlator in the low and mid frequency bands. The ASKAP observing bands are shifted by 48~MHz relative to the CRAFT bands (i.e., having central frequencies of 887.5~MHz and 1295.5~MHz, respectively), with a bandwidth of 288~MHz.

\subsection{CRAFT Data Products and Processing} \label{sec:software}

The CRAFT system described in \cite{Bannister565} was used to capture and download voltages from the desired beams for each source in this work. The subsequent offline correlation and calibration followed the procedure used for FRB localisations described in Section \ref{sec:methods} \citep[see also][]{Day2020}. As J1939 is the strongest of the four sources, for each beam and frequency band combination, one J1939 scan was designated as the `calibrator' scan while all other scans in each group were designated `target' scans. Prior to calibration, any radio frequency interference (RFI) present in the calibrator data was removed. The remaining clean portion of the data was used to derive phase and flux calibration solutions via the \textsc{aips} tasks \textsc{fring} and \textsc{calib}, which were used to correct for the frequency-dependent antenna-based delays, and \textsc{cpass}, which was used to determine the instrumental bandpass correction. Of note, the bandpasses for several of the antennas in the low-band data contained frequency-dependent gain features (e.g., dips or significant differences between the XX- and YY-polarisation product gains). However, these features, which can potentially be attributed to poor beam weights, did not prevent convergence on a good calibration solution, and so these data were included in the final sets processed for each beam. Along with this nominal calibration, in order to determine the affect of the ionosphere on the astrometric accuracy of the source positions given the $\sim$km baselines, a secondary calibrated dataset was obtained for each calibrator scan that further included ionospheric corrections derived with the \textsc{aips} task \textsc{tecor}. All solutions in each case were then applied to the target scans to form nominal and ionosphere-corrected datasets.

Stokes I (i.e., total intensity) images were then created using the \textsc{casa} task \textsc{tclean} in widefield, multi-frequency synthesis mode, forming a continuum image averaged across frequency for each combination of source, beam, central frequency, and calibration type. W-projection was used for the widefield deconvolution, with the individually calculated number of w-values being between $\sim 2 - 10$ planes. Due to several bright outlier field sources in the J1557 field, in addition to cleaning at the target source position, these outlier field sources were simultaneously cleaned using the `outlierfile' option in \textsc{tclean}. While this reduced the overall uncertainty in the fitted positions (such that these uncertainties were comparable to those obtained for the other sources), it had a negligible affect on the derived offsets, and since only marginal improvements were seen in the J1557 offsets, which are substantially larger than those derived for the other sources, this additional outlier imaging was deemed likewise unlikely to significantly alter the results obtained via the nominal imaging and so was not performed for the other sources.

Following the method used to astrometrically register image frames when localising FRBs (Section \ref{sec:methods}), the \textsc{aips} task \textsc{jmfit} was used to fit a 2-D Gaussian to a region of each snapshot image centred on the source and roughly equal to the size of the point spread function (PSF) to obtain the statistical position and uncertainty of each source in RA and Dec. While natural weighting results in the highest sensitivity and is generally used for the CRAFT imaging when obtaining continuum field source positions \citep[see, e.g.,][]{Day2020}, in the general radio image case, this is typically at the cost of resolution due to the potentially enlarged PSF and increased sidelobes. The difference between natural and uniform weighting, where the latter yields the highest resolution while sacrificing sensitivity, is expected to be relatively small for snapshot images \citep{briggs_thesis}. In order to determine the level of variation due to the weighting scheme in the fitted Gaussian, which depends on the PSF, two sets of images were made, with one using Briggs weighting with a robustness of 0.0 (i.e., halfway between uniform and natural) and the other using natural weighting (or equivalently Briggs weighting with a robustness of +2.0).

In order to investigate if the offsets exhibit a frequency dependence, the low-band data with the nominal calibration solutions applied were imaged as above (i.e., in multi-frequency synthesis mode) but in quarters of the band. We note that the centroid of the frequency-averaged image position in each sub-band then corresponds to the central frequency in each quarter. (See Section \ref{sec:200430} and Equation \ref{eq:cen_freq_def} therein for further details on the central frequency and its relation to the measured source centroid.) The low-band data were chosen since any frequency dependence in the offsets will be most prominent (and thus, more easily measured) at the lower frequencies. The source positions and uncertainties were then determined for each sub-band as described above.

\subsection{ASKAP Data Products and Processing} \label{sec:hardware}
A single, $\sim$5-minute ASKAP hardware correlator scan from each observation block was extracted and processed for each beam of interest and source. Basic flagging of the visibility data was performed to remove channels with known RFI or excessive circular polarisation; in the latter case, since none of the target sources are significantly circularly polarised, any excess is a result of RFI. While a regular bandpass observation with ASKAP includes 36 scans, with one scan per beam, only beams 15, 28, and 30 were extracted and used for the subsequent processing.

Scans for each source were split into separate measurement sets using the \textsc{casa} task \textsc{split}. As with the CRAFT software correlator data processing, a reference `calibrator' observation of J1939 was chosen for each beam such that the scan would correspond to the calibrator scan used in the CRAFT software correlator data processing, each of which have a reasonable elevation. A bandpass calibration was performed for the chosen calibrator visibility data measurement set of each of the three beams accounting for any slight pointing offsets specific to each beam. The resultant bandpass solution for each beam was then transferred to all scans captured for that beam. Similar to the low-band CRAFT software correlator data, features in the hardware correlator bandpasses (i.e., 4-MHz steps in the gains at the low end of the band) are indicative of either poor beamforming or issues with the On-Dish Calibrator\footnote{See memo ``017	The Utility of the ASKAP On-Dish Calibration System'' at https://www.atnf.csiro.au/projects/askap/ACES-memos} solutions. However, as these features remain unchanged throughout the observation run, they are not expected to significantly affect the astrometry.

Each calibrated scan was imaged using \textsc{tclean} in \textsc{casa}, with $2048 \times 2048$ 0.5-arcsec cells (that is, covering a field-of-view of $\sim0.3\degree$). The phase centre was set to the known RA and Dec.\ position of the source in each scan: that is, 19h39m25.0261s $-$63d42m45.625s for J1939, 15h57m51.4339s $-$00d01m50.413s for J1557, 22h14m38.5696s $-$38d35m45.009s for J2214 and 19h11m09.6528s $-$20d06m55.108s for J1911. The resultant images were then converted to \texttt{Miriad} \citep{Miriad1995ASPC...77..433S} images, and the task \textsc{imfit} was used to obtain a 2-D Gaussian fit of the source position within the central 10\% of the image.

\subsection{Effects of the Synthesised Beam on Fitting} \label{sec:psf_assumptions}

The estimated statistical positional uncertainties measured using \textsc{jmfit} are robust in the regime in which the PSF is well modelled by a Gaussian. Deviations from this can arise for arrays with sparse or clumpy ($u,v$) coverage, where the Gaussian that best fits the overall PSF may be narrower or broader than the central spike. If present, such a mismatch can lead to under- or over-estimated values for the statistical position uncertainty. See Section \ref{sec:field_src_offset_deriv} for further investigation into models that might be used to account for this.

Also of note, the output (i.e., positions and uncertainties) given by both \textsc{jmfit} and \textsc{imfit} are elliptical Gaussian approximations of the PSF. These are projected onto the RA and Dec.\ axes to obtain the positions and their uncertainties, and the position angles for these ellipses are derived. The direct use of these positions and uncertainties is appropriate in the case of a roughly circular synthesised beam, which is true for the majority of FRBs detected by ASKAP to date. However, when the ellipticity of the PSF is substantial (i.e., the case of a highly elongated beam), there is significant correlation between the measured uncertainties in RA and Dec., which is not captured by the direct use of the \textsc{jmfit} uncertainties. Thus, directly using these results would lead to a bias in the calculated mean offset and a misrepresentation of its uncertainty, an effect which worsens with decreasing frequency.

For unresolved sources, the uncertainty aligns with the PSF, but since there will be additional noise in each measurement, they will not all have the same position angle. Thus, there is no preferred axis on which to rotate the fitted Gaussian. For elongated beams, which can result from observations at very low elevations, a reference angle can be chosen. All fitted ellipses would then be rotated to this axis. The same would need to be done for the reference positions, and the comparison would be done in the rotated frame. The final results would then be re-projected onto the RA and Dec.\ axes. This process was used for FRB~181112 \citep{Prochaska231} and FRB~20201124A \citep{20201124a_low-band_ATel2021ATel14592....1D}, which were both observed at low elevations.

In this work, however, we assume a roughly circular beam for simplicity and note that this assumption will most significantly affect the low-elevation (e.g., J1557), low-frequency observations.

\subsection{Deriving Position Offsets and Dependencies} \label{sec:offset_unc_dep_derivation}

As per the method used for astrometric registration of FRB images, for each field source $i$ in the image, the fitted positions from the CRAFT and ASKAP data were compared to their catalogue counterpart positions to quantify the astrometric image-frame offsets in RA ($\alpha_i^e$) and Dec.\ ($\delta_i^e$), where $e$ denotes an estimated quantity, using the catalogue position as the reference (i.e., the CRAFT or ASKAP fitted position less the catalogue position). Since we have only a single source in the field for these observations and we assume a simple translation of the image frame (Section \ref{sec:methods}), these single-source measurements of the offsets and uncertainties directly correspond to the estimated mean position shift of the frame and the associated uncertainty in RA ($\mu^e_{\alpha}$, $\sigma^e_{\alpha}$) and Dec.\ ($\mu^e_{\delta}$, $\sigma^e_{\delta}$). This total offset uncertainty for each (RA, Dec.) pair was calculated by summing the reference and CRAFT or ASKAP uncertainties in quadrature. The selected sources are used as both ASKAP and Australia Telescope Compact Array (ATCA) calibrators and are thought not to possess significant frequency-dependent structure or structure on angular scales larger than the VLBI scales used to determine the ICRF3 positions. Nevertheless, we assume uncertainties of 10~mas in each coordinate of the catalogue position to account for such potential effects.

The RA and Dec.\ offsets obtained from the CRAFT and ASKAP data were compared against time, elevation, and angular separation for each `target' scan relative to the `calibrator' scan in order to constrain any dependencies on these parameters. Here, we have taken the Modified Julian Dates (MJDs) corresponding to the voltage dump triggers to be the times for each CRAFT scan and the scan start MJDs to be the times for each ASKAP scan. The time offsets for each scan were then calculated relative to the `calibrator' scan time (Figures \ref{fig:beam30_offvs.tel_highband} and \ref{fig:beam30_offvs.tel_lowband}). The scan elevations were derived using the above trigger or start times, the RA and Dec.\ coordinates of the beam centres, and the ASKAP latitude ($-26.697\degree$), longitude ($116.631\degree$ E), height above sea level ($361$~m), and radius from geocentre ($6374217$~m). The \verb|SkyCoord|, \verb|EarthLocation|, and \verb|AltAz| classes from the \verb|coordinates| subpackage of the \verb|astropy|\footnote{http://www.astropy.org} library \citep{astropy:2018} were used to obtain the RA and Dec.\ in degrees, to derive the location of ASKAP relative to geocentre, and to transform the source positions into an altitude and azimuth, respectively. We then take the altitude to be equivalent to the elevation. As with the time offsets, the elevation differences were calculated relative to the reference `calibrator' scans for each beam (Figures \ref{fig:beam30_offvs.tel_highband} and \ref{fig:beam30_offvs.tel_lowband}). The angular separations were likewise calculated for each source position relative to the `calibrator' scan's source position using the \verb|separation| task of the \verb|coordinates| subpackage.

As described in Section \ref{sec:software}, the possible frequency dependence in the CRAFT-derived offsets was also explored by sub-banding the low-band data and extracting the positions and uncertainties from each band (Figure \ref{fig:beam30_offvs.time_freq_lowband}). We then derive the source offsets for each scan and in each sub-band, and we fit the offsets versus wavelength for each scan in order to determine if the data are more consistent with a linear or non-linear dependence.

The CRAFT software correlator data are generally sensitivity limited due to the short ($\sim3$-second) integration available, as discussed in Section \ref{sec:methods}. The ASKAP hardware correlator, which runs continuously using the same input voltage data, should in principle produce identical results (modulo differences in the correlators, such as the geometric model used, signal quantisation, etc.) but with higher S/N due to the longer integration captured. In addition to differences in the correlated data products, including the potential effects of unmatched scan durations, residual offsets between the CRAFT and ASKAP-derived image frames can also result from the differing calibration solutions, data reduction strategies, and software packages (see \citealp[][]{Hotan2014PASA...31...41H,Hotan2021PASA...38....9H} for descriptions of the custom processing pipeline required for ASKAP given its simultaneous use of multiple PAFs [versus the single PAF beam used by CRAFT] to make large images using joint calibration and deconvolution). It is therefore of considerable interest to see how closely positions obtained from the ASKAP hardware correlator data products track those obtained from the CRAFT software correlator (Figures \ref{fig:beam30_offvs.tel_highband} and \ref{fig:beam30_offvs.tel_lowband}) since we would ideally use the ASKAP hardware correlator visibilities to derive calibration solutions with higher S/N. Accordingly, we derived residual offsets between the two frames by differencing positions derived from time-matched scans in the ASKAP hardware correlator data and the CRAFT software correlator data, and the individual positional uncertainties were added in quadrature to obtain the total uncertainty in these offsets.

Finally, we determined the total offset distributions for each beam. In general, these probability distribution functions (PDFs) are formed for RA and Dec.\ individually by summing over multiple Gaussian functions for which each estimated image-frame offset ($\mu^e_{\alpha}$ or $\mu^e_{\delta}$) and associated estimated uncertainty ($\sigma^e_{\alpha}$ or $\sigma^e_{\delta}$) pair is used as the mean and standard deviation. In order to account for the scan-specific mean offset imposed by our arbitrary selection of a given scan as the `calibrator', however, we re-reference the nominal PDF such that we obtain a `true' distribution -- i.e., the distribution that would result from using each scan as the `calibrator' scan exactly once. This is accomplished by looping over the offset and uncertainty pairs and taking the difference between all offsets and uncertainties and each pair in turn, forming two matrices with dimensions given by the number of pairs (i.e., for the offset matrix, $\rm N_{offset} \times N_{offset}$, where $\rm N_{offset}$ is the number of offsets in a given beam, and similarly for the uncertainty matrix). A new set of Gaussian distributions was then evaluated using these re-referenced offset and uncertainty values as the mean and standard deviation inputs, and a total PDF was obtained for each beam by summing over these Gaussian distributions and normalising by the number of input PDFs (Figures \ref{fig:hw_applied} and \ref{fig:tru_dist}).

\section{Results and Analysis} \label{sec:res_analysis}

As discussed in Section \ref{sec:methods}, the current method of registering the CRAFT reference frame to that of the ICRF3 (i.e., estimating the overall systematic shift between the frames in RA and Dec.) uses a comparison between the continuum background sources detected in the field image and the counterpart source positions obtained from observations with higher confidence calibration solutions. The degree of any systematic shift present in the reference frame of the image and the level of source smearing due to residual phase errors are dependent on how accurately the calibration solutions can be interpolated across time and space. If quantified and completely corrected for in the manner described in Section \ref{sec:methods}, the systematic shift is of no concern. However, our ability to measure this shift is limited by the number of field sources detected and their S/N.

Given these limitations, we conducted two investigations. First, in order to characterise the typical astrometric accuracy in a range of observational circumstances, we examined a set of potential sources of systematic error that were thought most likely to affect the quality of the interpolated calibration solutions and thereby the final astrometric accuracy we are able to obtain (Section \ref{sec:offset_deps}). Second, as described in Section \ref{sec:methods}, we explored the feasibility of applying the hardware-correlator-derived calibration solutions to the CRAFT software correlated data as a means of reducing the S/N limitation and thereby improving the overall accuracy of the final FRB position (Section \ref{sec:future_obs}).

\subsection{Offset Dependencies} \label{sec:offset_deps}

As described in Section \ref{sec:offset_unc_dep_derivation}, we compared the positional offsets derived for each target source scan with the relative separation between the target scan and the selected calibrator scan in time, elevation, and angular distance to determine any potential dependence on these observational factors. This was done for each beam and both frequency bands. Since no significant differences were found between the offset distributions for the beams (see Section \ref{sec:ast_vs_FRBdist} and Figure \ref{fig:tru_dist}) and the offset dependencies for each beam were consistent, we take beam 30 to be a representative beam. We find no significant trend in offset versus angular separation from the calibrator scan and so omit these plots.

\begin{figure*}
\begin{center}
\includegraphics[width=0.7\textwidth]{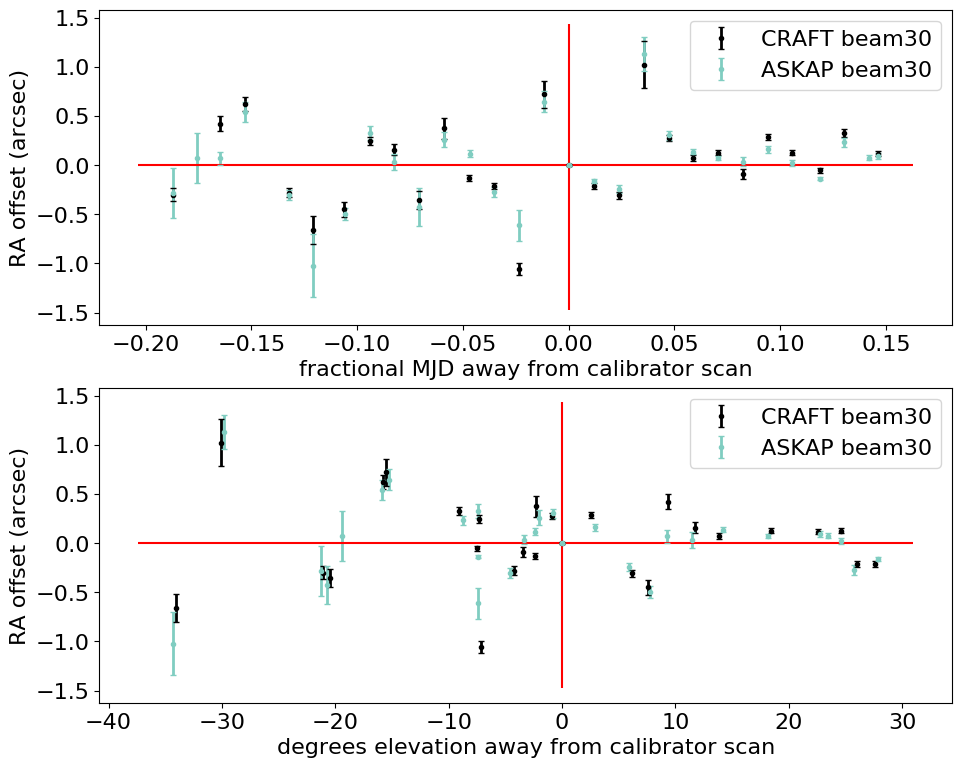}
\includegraphics[width=0.7\textwidth]{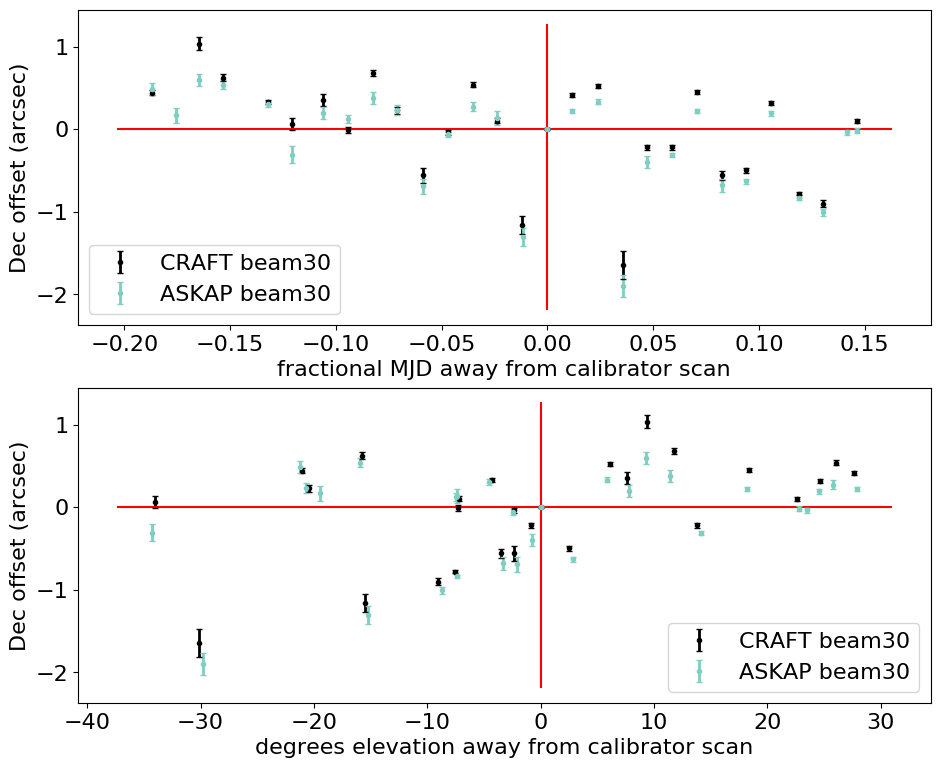}
\caption{Mid-band positional offset dependencies on time and elevation. Panels 1 and 3 show the RA and Dec.\ offsets for beam 30 versus the fraction of the MJD relative to the calibration scan MJD. Panels 2 and 4 show these beam 30 offsets against the differential elevation relative to the calibrator scan. The corresponding offset dependencies on time and elevation for the beam 15 and beam 28 data are comparable, and so only beam 30 is shown. The red lines mark the zero-offset in position and zero-offset from the calibrator scan in either time or elevation.}
 \label{fig:beam30_offvs.tel_highband}
\end{center}
\end{figure*}

\begin{figure*}
\begin{center}
\includegraphics[width=0.7\textwidth]{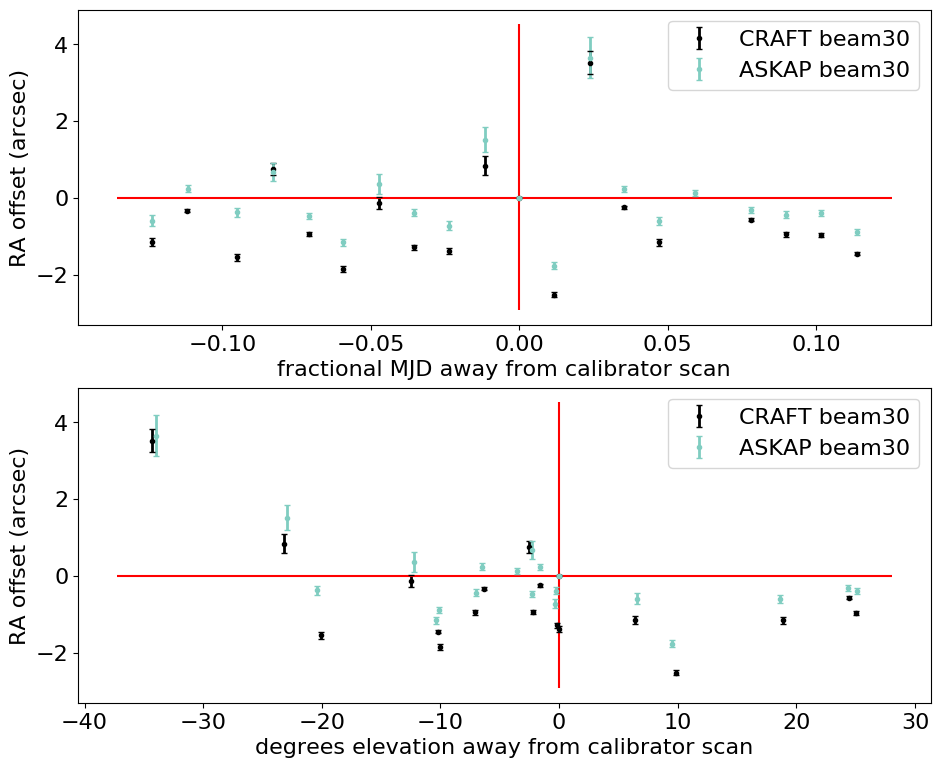}
\includegraphics[width=0.7\textwidth]{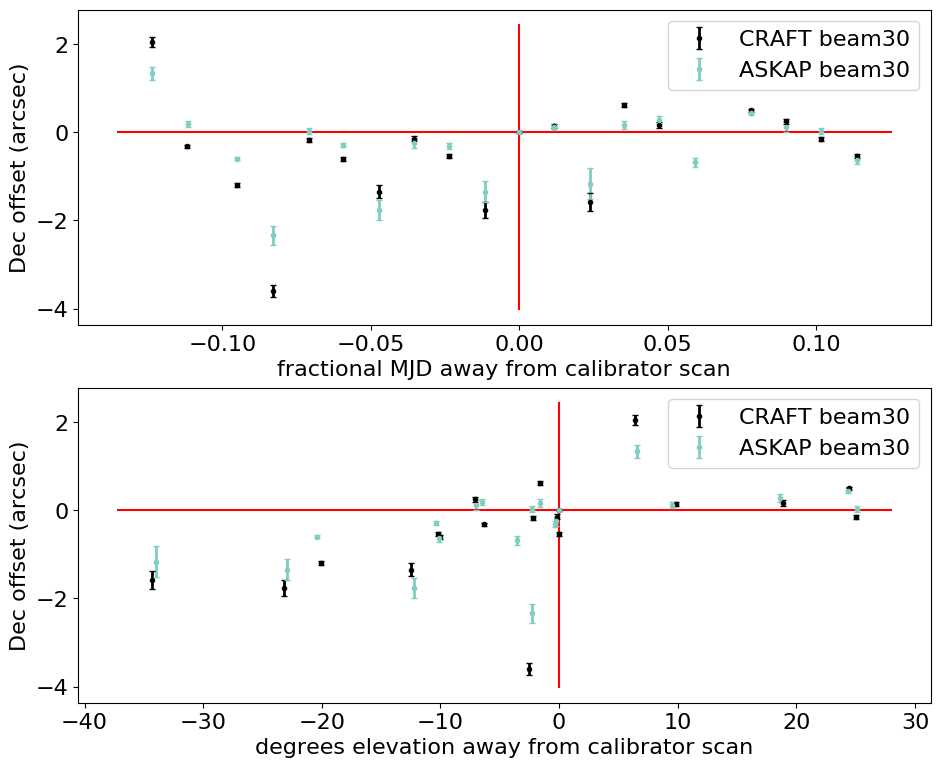}
\caption{Same as Figure \ref{fig:beam30_offvs.tel_highband} for the low-band positional offset dependencies on time and elevation. As with the mid-band offset dependencies, the overall structure of the beam 30 trends are comparable to those seen in beam 15 and beam 28.
In contrast to the mid-band results, the RA and Dec.\ offset dependencies on time and elevation separation from the calibrator scan in the low-band data are more pronounced. This is due in part to the larger beam size. Notably, the five points in each panel that have the largest offsets and uncertainties are from J1557. Here, the hardware and CRAFT offsets do not have a consistent average differential offset from each other, in contrast to the mid-band data.}
 \label{fig:beam30_offvs.tel_lowband}
\end{center}
\end{figure*}

Figures \ref{fig:beam30_offvs.tel_highband} and \ref{fig:beam30_offvs.tel_lowband} respectively show the mid- and low-band offset dependencies on the fractional separation in time (MJD) and elevation (degrees) relative to the calibrator scan for RA (top two panels) and Dec.\ (bottom two panels). We find some dependence on time and elevation in both bands, with the trends in Dec.\ more pronounced than those in RA, but these are generally weak in both directions with a lot of scatter.

This scatter is more significant for the low-band data. However, the frequency-dependent gain features in the bandpasses discussed in Section \ref{sec:software} possibly contribute to this increased scatter. While good beam weights will lead to a well-behaved PAF beam that closely approximates the desired Gaussian form, poor beam weights could potentially cause deviations from this ideal in a frequency dependent way. If the bandpass is then taken at a fixed location (e.g., the nominal beam centre), the resulting gain as a function of frequency will be distorted (relative to that obtained from a more Gaussian PAF beam), which is true of many of the low-band scans. However, as this is true of both the data presented here and other typical ASKAP observations, our interpretation should be valid for real-world observations in general.

Notably, the largest offsets in elevation for the low-band data are those of J1557, which is the most northerly source in our sample. As discussed in Section \ref{sec:psf_assumptions}, our positional fitting process neglects correlations between the right ascension and declination uncertainties. At low elevations, however, the synthesised beam becomes increasingly elongated, leading to both a larger major axis for the synthesised beam and an increasing covariance between the errors in these two coordinates (depending on the synthesised beam's position angle). However, the dependence of the potential underestimation of offset on elevation due to the changing synthesised beam properties is expected to be smaller than the overall trend seen here, and so we conclude that there is some offset dependence on elevation.

As noted in Section \ref{sec:software}, we also imaged the low- and mid-band visibilities using Briggs weighting with a robustness of 0.0. We detected no significant deviations in the offset trends (with time, elevation, or angular separation) found when using the Briggs versus naturally weighted images for either frequency band.

\begin{figure*}
\begin{center}
\includegraphics[width=0.7\textwidth]{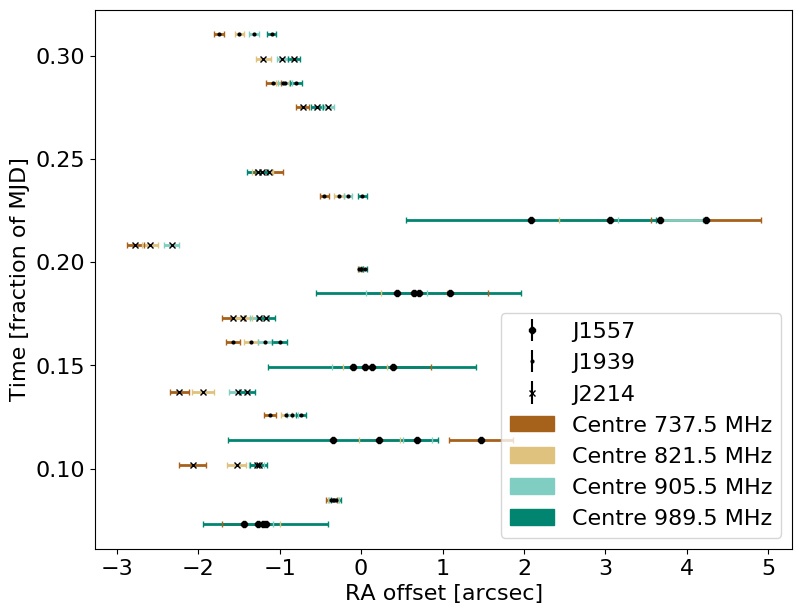}
\includegraphics[width=0.7\textwidth]{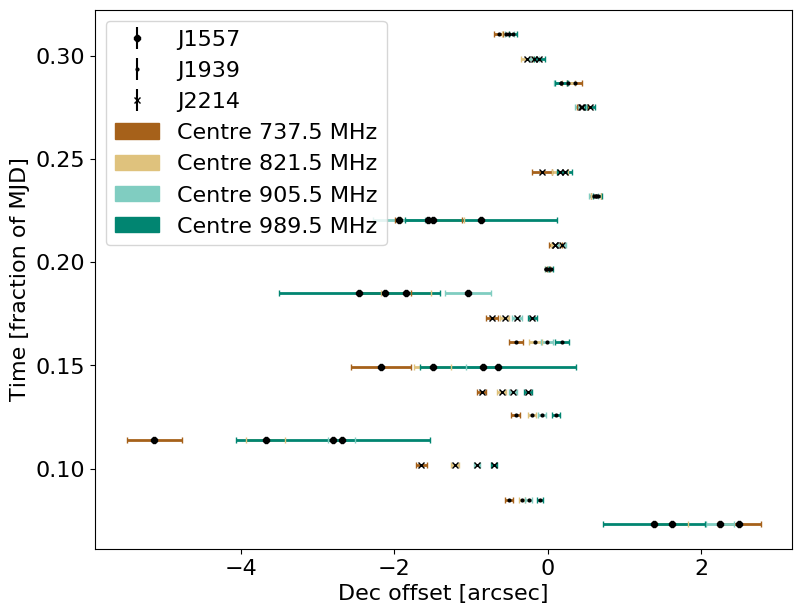}
\caption{RA (top) and Dec.\ (bottom) beam 30 offsets derived from the sub-banded data versus fraction of the MJD (i.e., the time for each scan of the CRAFT data). We found no substantial differences in the overall trend for each beam, and so we take beam 30 to be representative. Colour represents the central frequency of the four sub-bands, while the sources are distinguished by marker style. Overall, the offsets get smaller with increased frequency.}
 \label{fig:beam30_offvs.time_freq_lowband}
\end{center}
\end{figure*}

Figure \ref{fig:beam30_offvs.time_freq_lowband} shows the frequency dependence in the beam 30 offsets versus time. As with the offset distributions, we found no significant differences between the beams and therefore take beam 30 to be representative. While the dependence appears to be non-linear in some scans, fitting the offsets versus wavelength in each scan showed that most scans are adequately described by a linear fit, with only a few scans (across all beams) being more consistent with a nonlinear model. Linear growth in offset with wavelength is consistent with a frequency-independent phase error. In particular, the size of the synthesised beam grows linearly with wavelength, and so given a fixed fraction of the PSF (i.e., a constant phase error), the offsets would likewise grow linearly with wavelength.

\subsection{Implications for Future Observations} \label{sec:future_obs}

\begin{table*}
\caption{The 16th, 50th, and 84th percentiles of the CRAFT-ASKAP cumulative distribution functions for the low- and mid-band observations. Note that the top set of values were derived from the offset distributions made using naturally weighted CRAFT images while the bottom set are those from Briggs weighted CRAFT images.}
\centering
\begin{tabular}{c|cc|cc}
\hline\hline
Percentile & low-band RA & low-band Dec. & mid-band RA & mid-band Dec.  \\
 & (arcsec) & (arcsec) & (arcsec) & (arcsec) \\
\hline%
  16 & $-0.610$ & $-0.49$ & $-0.088$ & $-0.02$ \\
  50 & 0.017 & $-0.08$ & 0.070 & 0.11 \\
  84 & 0.463 & 0.18 & 0.222 & 0.26 \\
\hline
  16 & $-0.475$ & $-0.39$ & $-0.165$ & $-0.01$ \\
  50 & 0.017 & $-0.06$ & $-0.00048$ & 0.11 \\
  84 & 0.375 & 0.19 & 0.123 & 0.23 \\
\hline\hline
\end{tabular}
\label{tab:percentiles}
\end{table*}

Fundamentally, the method used to obtain astrometric corrections for the ASKAP frame requires the creation of a model of the field and the use of this model to determine the positional corrections to be applied. This can be accomplished either through self-calibration to a sky model formed from the data or via field source comparison to an external model known to have sufficient accuracy (i.e., the current method in use). In the case of the former, this requires a reasonably high astrometric registration accuracy. At present, the ASKAP hardware correlator data when fully processed has a known systematic astrometric offset of up to $\sim 1$~arcsec, which is well above the statistical uncertainty in the position of a typical FRB ($\sim 100$~mas). However, this is expected to improve in the future, with a reasonable estimate of the accuracy limit attainable likely on the order of 0.05~arcsec -- i.e., well within the uncertainty obtainable for a high S/N FRB and comparable to that of the Faint Images of the Radio Sky at Twenty-Centimeters (FIRST) catalogue \citep{Becker_FIRSTsurvey1995ApJ...450..559B}.

As discussed in Section \ref{sec:methods}, the current method is largely limited by the number and brightness of field sources present, which vary stochastically from field to field. To that end, when employing this comparison method, improvements to the typical accuracy we can obtain in any given field using the CRAFT software correlator data must come from an increase in sensitivity, which is attainable via the longer integration times used for the ASKAP hardware correlator data. For example, a 60-s integration (i.e., 20x that of CRAFT) would result in $\sqrt{20} \sim 5$x higher sensitivity than the current CRAFT data products and would therefore reduce the uncertainty in a typical field to that below the statistical positional uncertainty for a typical ASKAP FRB. While the integration time of the ASKAP hardware correlator data (default 10 seconds) would depend on the configuration required for the observation with which CRAFT would run commensally, we would be able to reprocess a subset of the data suitable for use with the CRAFT voltages, including selecting a reasonable duration (i.e., longer slices of data for faint fields if needed) roughly centred on the temporal position of the FRB. The use of 5-minute scans in this work and the results obtained are therefore representative of what would ultimately be used and the typical corrections these scans would yield when using the hardware correlator data to conduct the field source comparison.

However, directly applying corrections derived using the hardware correlator data products to the CRAFT data products is only feasible if there are no systematic differences resulting from the different data paths. These could, for instance, result from differences in the geometric models used in the two correlators, the effects of the requantisation of the CRAFT voltages, the differing calibration solutions, or differences in how the data are processed. The datasets we present here allow us to place upper limits on the maximum size of any such systematic differences.

\begin{figure*}
\begin{center}
\includegraphics[width=0.85\textwidth]{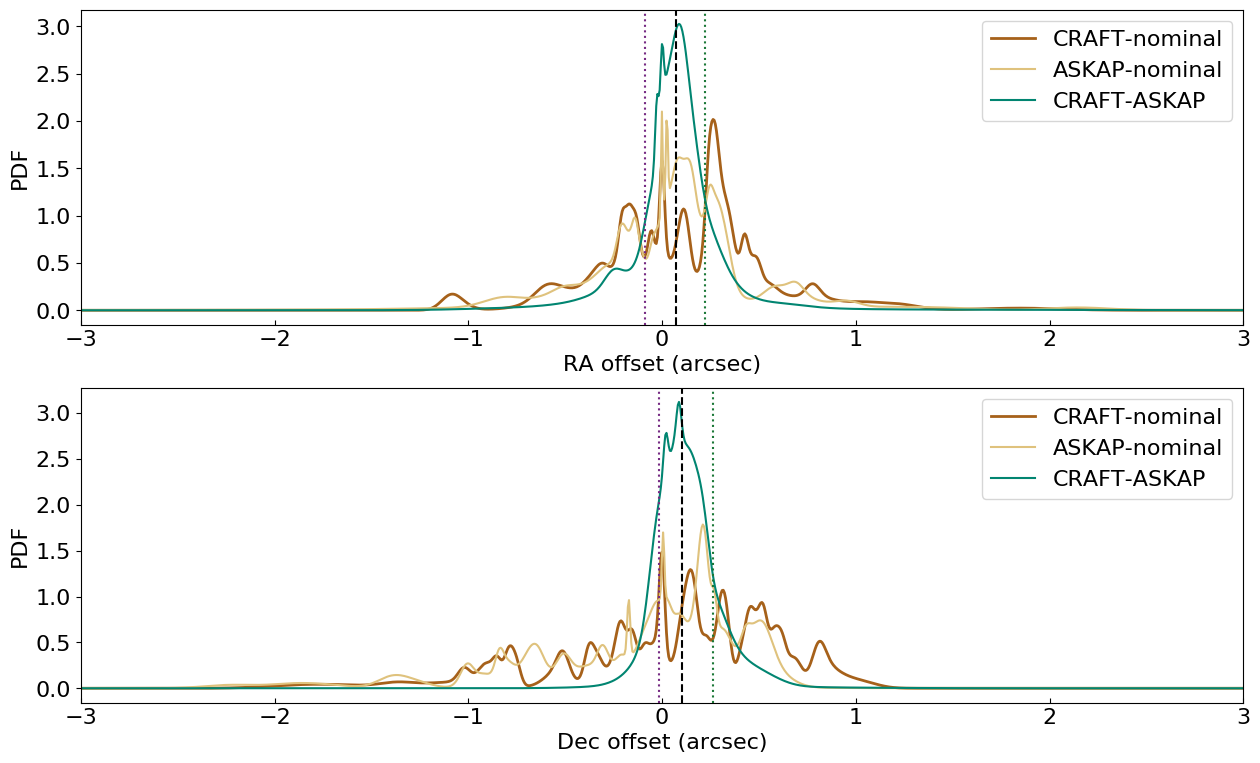}
\includegraphics[width=0.85\textwidth]{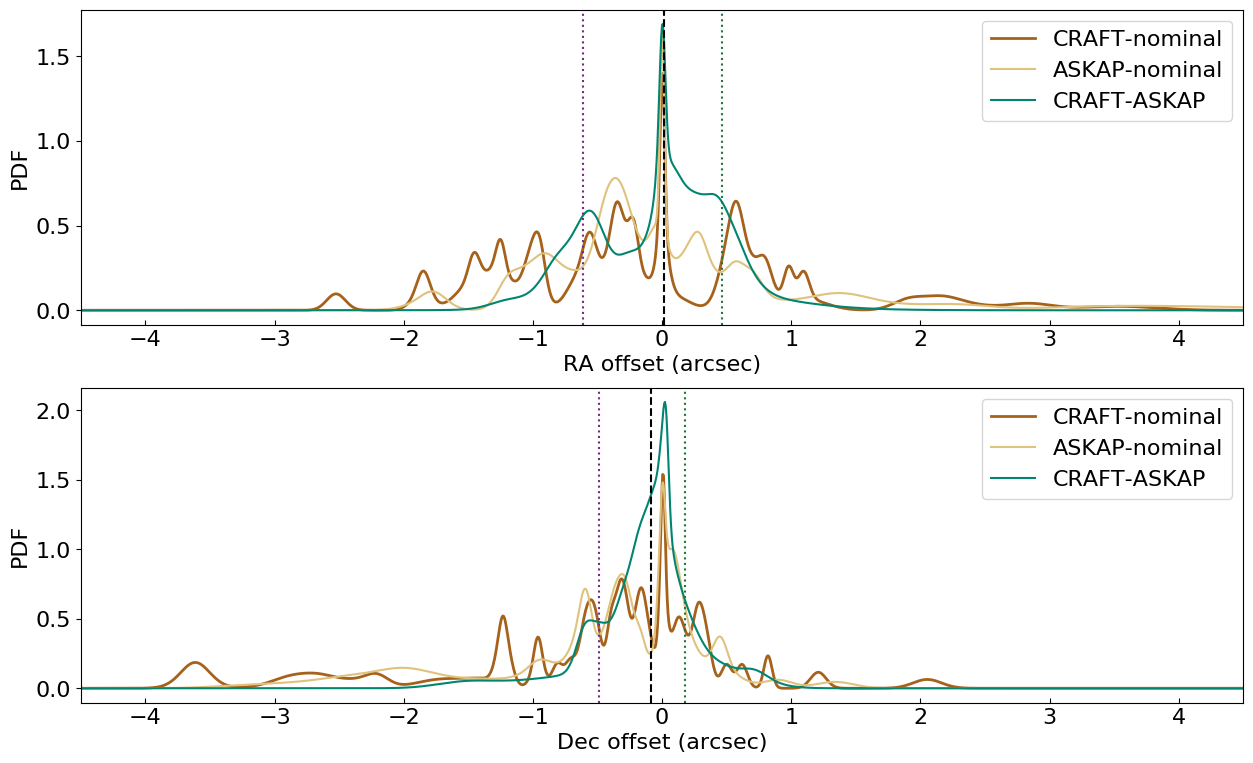}
\caption{RA and Dec.\ offset probability density functions marginalised over the three beams for the mid-frequency band (top two panels) and low-frequency band (bottom two panels). The `CRAFT-nominal' and `ASKAP-nominal' are respectively the PDFs formed from the CRAFT software correlator and the ASKAP hardware correlator positions less the nominal source positions, and the `CRAFT-ASKAP' is the PDF formed from the CRAFT software correlator positions less the ASKAP hardware correlator positions. Also shown are the median (black dashed line) and the 16th (purple dotted line) and 84th (green dotted line) percentiles (together the 68\% confidence limits) of the `CRAFT-HW' cumulative distribution function.}
 \label{fig:hw_applied}
\end{center}
\end{figure*}

Figure \ref{fig:hw_applied} shows the offset probability distribution functions (marginalised over beam) obtained for both the CRAFT software correlator and ASKAP hardware correlator positions (with the former measured from the naturally weighted images) less the catalogue positions. The PDF formed from the difference of the CRAFT- and ASKAP-derived positions averaged over beam is also shown along with the 16th, 50th, and 84th cumulative percentiles derived from evaluating the normalised cumulative distribution function of this difference (see also Table \ref{tab:percentiles}).

The mean difference between the offsets derived using the CRAFT and ASKAP correlators should be zero in the case of identical inputs, centre times, calibration solutions, and geometric models. However, the inputs are not identical (e.g., the ASKAP scans are much longer than the CRAFT ones), the times of both scans are not precisely centred (see, e.g., Figures \ref{fig:beam30_offvs.tel_highband} and \ref{fig:beam30_offvs.tel_lowband}), the derived calibration solutions (in this simple case of bandpass and phase calibration only) differ (e.g., due to the larger CRAFT bandpass, which is also not centred at the ASKAP band centre), and there are potentially small differences in the geometric models used. These relative deviations can lead to a nonzero mean difference between the positions that is expected to change with each observation, as evidenced by the differential offset between the positional offsets derived for each scan (see Figures \ref{fig:beam30_offvs.tel_highband} and \ref{fig:beam30_offvs.tel_lowband}). Any mean measurement, then, is a function of the sources and the parameter space sampled. In order to sample this parameter space in a representative manner, we use a set of sources with a range of RA and Dec.\ positions and relative separations in time, elevation, and angular offset. For each of the frequency bands and the combined positional axes, we find that the central 68\% of the sample spans the differential offset range of $\sim 0.5 - 0.6$~arcsec (low-band) and $\sim 0.2 - 0.3$~arcsec (mid-band) (i.e., taking the maximum absolute values of the 16th and 84th percentiles across position shown in Table \ref{tab:percentiles}). Given the distributions are not perfectly Gaussian, we conservatively estimate the residual systematic offset between the ASKAP and CRAFT frames to be the larger of the position-combined asymmetric percentiles. Accordingly, in the simple case of applying a bandpass and phase calibration and in the limit of high S/N (which will always be the case with the hardware correlator data), we estimate the systematic uncertainty of low-band and mid-band observations, respectively, when using hardware correlator-derived corrections will be $\sim 0.6$~arcsec and $\sim 0.3$~arcsec in RA and Dec.

Additionally, as shown in Figures \ref{fig:beam30_offvs.tel_highband} and \ref{fig:beam30_offvs.tel_lowband}, although the residual offsets between positions derived from the hardware and software correlator data products are not constant, there is no dependence on time or elevation in the systematic offset between the two frames. Likewise, we find no trend when comparing the offsets versus angular separation. We therefore conclude that we should be able to obtain good solutions when applying the hardware correlator-derived offsets to the software correlator data products regardless of differences in time, elevation, or angular separation.

As discussed in Section \ref{sec:software}, in addition to natural weighting, images were made using Briggs weighting with a robustness of 0.0 to quantify any variation or improvement in the offsets and their uncertainties due to the resultant increase in resolution. In the high-S/N regime, for reasonable ($u,v$) coverage, and across a wide range of elevations, we find that Briggs weighting with a robustness of 0.0 yields improvement of 17\% and 33\% in the 68\% confidence intervals we derive respectively for the low- and mid-band residual offsets between the CRAFT and ASKAP image frames (Table \ref{tab:percentiles}). This is unsurprising in the high-S/N case we've studied here since Briggs weighting will result in a closer approximation of the PSF when fitting the positions (Section \ref{sec:psf_assumptions}). However, while Briggs weighting performs better for the parameter space we've tested here, future investigations should confirm that this result holds in the low-S/N regime as well as for observations with different antenna arrangements or smaller sub-arrays. Typically, CRAFT field image sources have low S/N, and so both the loss of sensitivity and higher resolution obtained when using Briggs weighting could result in reduced S/N and poorer approximations of the true PSF. Of note, once the use of the hardware correlator data is employed, these data will always be in the high-S/N regime, mitigating any issues arising from low S/N sources. Further studies on the affects of array size and configuration as well as fitting low-S/N sources on the typical offsets and uncertainties will be conducted in a future work.

\subsection{Modelling Large-Scale Ionospheric Effects} \label{sec:ionocal}

As described in Section \ref{sec:software}, along with the datasets obtained by applying the nominal calibration solutions, we also produced datasets which additionally model the variations in the ionosphere -- in particular, the dispersive delays caused by deviations in the total electron content (TEC) between sightlines -- by including corrections derived using the \textsc{aips} task \textsc{tecor}. For this, we used the International GPS Service for Geodynamics (IGS) Global (IGSG) ionosphere maps in the The IONosphere Map EXchange (IONEX) format for each observing day to derive these solutions\footnote{The IONEX files are available from https://cddis.nasa.gov/}.

We found that the day the mid-band data were recorded had increased ionospheric activity relative to the day the low-band observations were conducted. This led to larger ionosphere corrections for the mid-band data. To determine the overall affect of these solutions, we differenced the offsets calculated for each scan, using the offsets derived from the data with the nominal calibration solutions applied as the reference. For the low-band data, we find typical differential offsets for beams 15, 28, and 30 in RA and Dec.\ of (0~mas, 0~mas), (0~mas, 0~mas), and (12~mas, 10~mas), with differential offsets up to 13~mas in RA across all beams and 10, 20, and 30~mas in Dec.\ for beams 15, 28, and 30, respectively. In contrast, for the mid-band data, we find differential offsets of (42~mas, 60~mas), (41~ms, 50~mas), and (41~ms, 60~mas) in RA and Dec.\ for beams 15, 28, and 30, respectively, with maximum values of 985, 938, and 879~mas in RA and 350, 320, and 290~mas in Dec.\ respectively for beams 15, 28, and 30.

Since the corrections in the model used are smoothed over approximately two hours and roughly $2\degree$, this model is not well-suited to small arrays due to this coarse sampling and the need to interpolate over a much larger spatial extent than the resolution of the array. This method therefore does not probe small-scale ionospheric effects but does provide an estimate of the large-scale effects of the ionosphere over a range of activity levels. Our results indicate that the ionosphere might contribute to the spatial and temporal offsets we see, but further investigation is required.

\section{Comparison with the FRB Offset Distribution} \label{sec:ast_vs_FRBdist}

Together with characterising the typical offset distributions expected in the CRAFT and ASKAP positional frames and any dependence on observational parameters (Section \ref{sec:res_analysis}), we also wish to establish how well our offset distributions measured in this work match the published FRB offset distributions and evaluate if the method currently used to derive the field source offsets and uncertainties is optimal (see Section \ref{sec:field_src_offset_deriv} for the latter).

\begin{figure*}
\begin{center}
\includegraphics[width=0.75\textwidth]{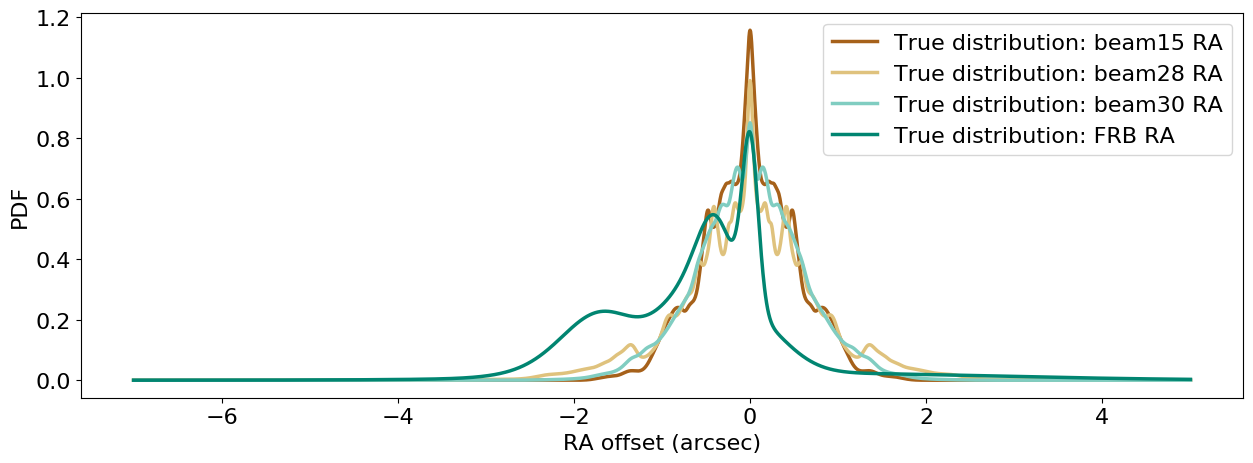}
\includegraphics[width=0.75\textwidth]{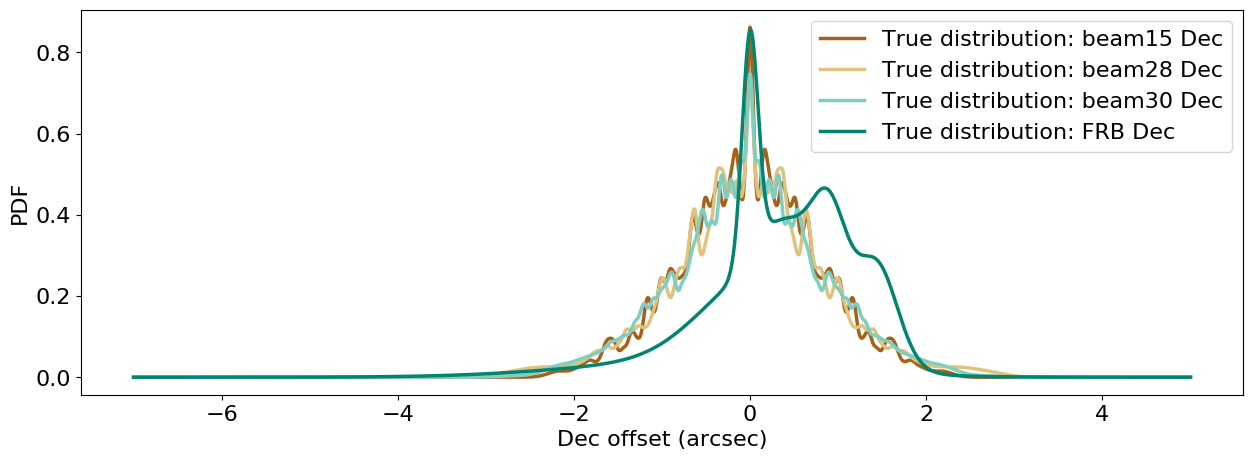}
\includegraphics[width=0.75\textwidth]{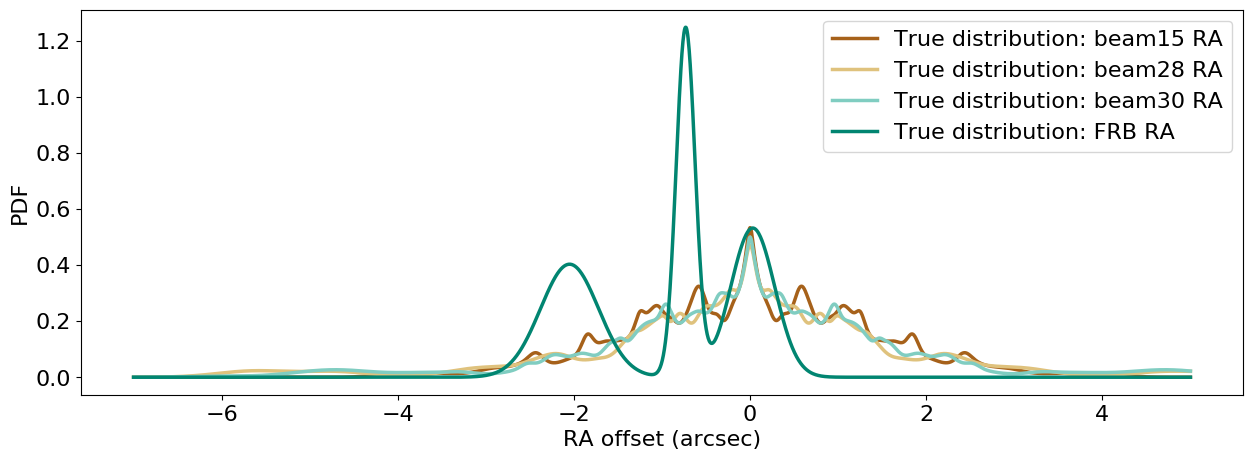}
\includegraphics[width=0.75\textwidth]{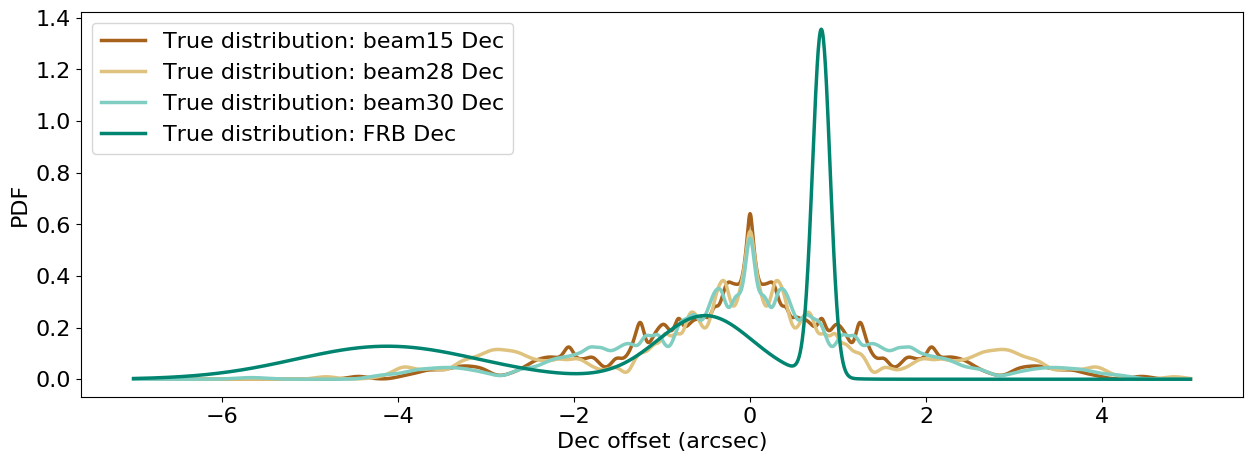}
\caption{Re-referenced probability distribution functions for the beam 15, 28, and 30 astrometric offsets for the mid-band data (top two panels) and low-band data (bottom two panels) imaged using natural weighting with the FRB offset distributions shown for comparison. The offset distributions obtained for the strong point sources are both consistent with each other and largely consistent with the FRB offset distributions obtained using the published offsets and uncertainties for the mid- and low-band detected FRBs, respectively. Note that the low-band FRB PDF was formed with only 3 FRBs, while the mid-band distribution was formed using 8 FRBs.}
 \label{fig:tru_dist}
\end{center}
\end{figure*}

Figure \ref{fig:tru_dist} shows the `true' (i.e., re-referenced and combined) offset distributions of both RA and Dec.\ in each beam (15, 28, and 30) for both frequency bands, as described in Section \ref{sec:offset_unc_dep_derivation}, along with the FRB offset distributions for each case. The mid-band FRB offset PDF comprises 8 FRBs, while the low-band PDF was formed using only 3 FRBs (Table \ref{tab:frbs}). Since the PDFs are formed by summing the individual Gaussian distributions evaluated using the offset and final uncertainty derived for the individual FRBs (as detailed in Section \ref{sec:offset_unc_dep_derivation}), the trimodal PDF in the low-band case is the result of the small number of FRBs available in this band, whereas the greater number of mid-band FRBs forms an overall smoother summed distribution.

We find that the RA and Dec.\ offset distributions measured in each beam are both highly consistent with each other and overall consistent with the published FRB offset distributions in each direction and frequency band (Figure \ref{fig:tru_dist}). For the data presented in this work, the observation of fewer scans in the low-band and the increased number of large offsets results in an overall broadened distribution when compared to the mid-band offsets. Of note, the largest offsets in the low-band FRB PDFs are those of FRB~200430, which has a known frequency-dependent offset in Dec., resulting in a substantially larger Dec.\ offset than the other FRBs (in either frequency band) and a consequently broadened uncertainty range (see Section \ref{sec:200430}); the offset distribution measured for FRB~200430 is nevertheless consistent within its $1\sigma$ uncertainty region and that of the data in this work. As with the offset PDFs derived for the strong calibrator sources presented here, the FRB offset distributions are broader at lower frequencies. Given the degree of consistency between the offset distributions derived from the data described in this work and the FRB sample, therefore, we conclude that the measured offsets and uncertainties of the published FRBs are consistent with expectations based on this work.

\begin{table*}
\caption{FRBs used to form the low- and mid-band distributions, as indicated in the $\nu_{\rm obs}$ column, shown in Figure \ref{fig:tru_dist}. A subset of these FRBs was also used in the analysis detailed in Section \ref{sec:field_src_offset_deriv}. Where this is the case, we list the number of field sources used for a given FRB ($\rm{N_{src}}$), the number of degrees of freedom (NDF), the variance ($s^2$), and the one-sided p-value from the $\chi^2$ test. The total values for an overall test using all FRBs are given in the last row, with the total variance given by Equation \ref{eq:tot_var}.}
\centering
\begin{threeparttable}
\begin{tabular}{lllllll}
\hline\hline
Source & $\nu_{\rm obs}$ & $N_{src}$ & NDF & $s^2$ & $p$ & Reference  \\
\hline%
  FRB~180924 & mid band & & & & & \cite{Macquart2020_DMz}\tnote{\ddag} \\
  FRB~181112 & mid band & & & & & \cite{Prochaska231} \\
  FRB~190102 & mid band & 2 & 2 & 1.79 & 0.41 & \cite{Macquart2020_DMz} \\
  FRB~190608 & mid band & 3 & 4 & 6.90 & 0.14 & \cite{Day2020}\tnote{$\dagger$} \\
  FRB~190611 & mid band & 2 & 2 & 1.02 & 0.60 & \cite{Macquart2020_DMz} \\
  FRB~190711 & mid band & 4 & 6 & 5.41 & 0.49 & \cite{Day2020}\tnote{$\dagger$} \\
  FRB~190714 & mid band & 3 & 4 & 6.50 & 0.16 & \cite{Heintz_hostII_2020ApJ...903..152H} \\
  FRB~191001 & low band & 3 & 4 & 10.87 & 0.028 & \cite{Bhandari2020ApJ...901L..20B} \\
  FRB~191228 & mid band & 2 & 2 & 0.68 & 0.713 & Bhandari et al. (in prep) \\
  FRB~200430 & low band & 8 & 14 & 69.66 & $2.22 \times 10^{-9}$ & \cite{Heintz_hostII_2020ApJ...903..152H} \\
  FRB~200906 & low band & 7 & 12 & 50.24 & $1.27 \times 10^{-6}$  & Bhandari et al. (in prep) \\
  \hline
  Totals: & & 32 & 48 & 152.40 & $2.19 \times 10^{-12}$ & \\
\hline\hline
\end{tabular}
\begin{tablenotes}[flushleft]
        \item[$\ddag$] Position originally reported in \cite{Bannister565} and updated in \cite{Macquart2020_DMz}.
        \item[$\dagger$] Position originally reported in \cite{Macquart2020_DMz} and updated in \cite{Day2020}.
    \end{tablenotes}
\end{threeparttable}
\label{tab:frbs}
\end{table*}

\subsection{Optimising Field Source Offset Derivation} \label{sec:field_src_offset_deriv}

As described in Section \ref{sec:methods}, the current method used to correct the astrometry of the FRB image frame uses a simple weighted mean to derive the final mean image offsets in RA and Dec.\ (i.e., assuming any offsets to be simple translations of the image frame) along with the associated systematic uncertainties (i.e., either the error in the weighted mean of the estimated uncertainties or the scatter in the points about the mean in the case of scatter-dominated offsets, as discussed in Section \ref{sec:methods}). We note that, as we have done in this work (Section \ref{sec:psf_assumptions}), this method uses positional uncertainties projected onto RA and Dec., which loses the ability to show covariances in the final systematic uncertainty between RA and Dec. In order to investigate if the current method is optimal and, if not, what a preferred model might be, we test various hypotheses to determine the most reasonable estimates of the true mean FRB image frame offsets ($\mu^t_{\alpha}$, $\mu^t_{\delta}$) and uncertainties ($\sigma^t_{\alpha}$ and $\sigma^t_{\delta}$). We use the $t$ and $e$ superscript notation throughout to respectively indicate the true and estimated quantities.

For each FRB, the data consist of $\rm N_{src}$ field source offsets in RA, $\alpha^e_i$, and Dec., $\delta^e_i$ (i.e., taking the relative offsets of the $i^{th}$ field source from the nominal source positions as estimates of the image frame offsets in RA and Dec.) and estimates of the uncertainties in these individual offsets ($\sigma^e_{\alpha,i}$, $\sigma^e_{\delta,i}$). Table \ref{tab:frbs} lists the FRBs used for this analysis. FRB~180924 and FRB~181112 were not included in the sample because their field source comparisons were based on a single continuum source in their respective fields.

Our initial hypothesis ($\rm H_0$) assumes the provided estimated uncertainties ($\sigma^e_{\alpha,i}$, $\sigma^e_{\delta,i}$) on the measured field source offsets correctly estimate the true uncertainties in the image frame offsets as measured by each source. We also assume that all estimated uncertainties of the measured field source offsets are independent Gaussian random variables with a mean of zero and the stated deviation. In this case, the $i^{\rm th}$ field source for some FRB with RA and Dec.\ offsets [$\alpha^e_i$, $\delta^e_i$] and estimated errors [$\sigma^e_{\alpha,i}$, $\sigma^e_{\delta,i}$] will be related to the true mean image offsets [$\mu^t_{\alpha}$, $\mu^t_{\delta}$] via
\begin{align}
    [\alpha,\delta]^e_i &= \mu^t_{[\alpha,\delta]} + d[\alpha,\delta]^e_i \\
    d[\alpha,\delta]^e_i &\sim N(0,\sigma^e_{[\alpha,\delta],i}), \label{eq:rand_deviate}
\end{align}
\noindent where $N$ denotes the Normal distribution and we use the square bracket notation throughout to indicate evaluation of equations using either RA or Dec.\ values.

Under the assumption that $\rm H_0$ is true, the best estimates $\Delta[\alpha,\delta]$ of the true mean image offsets $\mu^t_{[\alpha,\delta]}$ are given by the weighted estimates
\begin{equation} \label{eq:estimates}
    \begin{aligned}
        \Delta[\alpha,\delta] &= \frac{\sum_i w_i [\alpha,\delta]^e_i}{\sum_i w_i} \\
        w_i &= \frac{1}{\Big({\sigma^e_{[\alpha,\delta],i}}\Big)^2},
    \end{aligned}
\end{equation}
\noindent where $w_i$ is the weight for either the RA or Dec.\ of the $i^{th}$ source. To test the validity of our hypothesis, we can use a chi-squared ($\chi^2$) test to construct an unbiased estimator of the sample variance given by
\begin{equation} \label{eq:variance}
    s^2_{[\alpha,\delta]} = \sum w_i ([\alpha,\delta]^e_i - \Delta[\alpha,\delta])^2,
\end{equation}
\noindent since $s^2_{[\alpha,\delta]} \sim \chi^2_{\rm{N_{src}}-1}$, where $\chi^2_{\rm{N_{src}}-1}$ is a $\chi^2$ distribution with ${\rm{N_{src}}-1}$ degrees of freedom. Simultaneously checking both RA and Dec.\ yields
\begin{equation} \label{eq:ra_dec_variance}
    \begin{aligned}
        s^2_{\alpha,\delta} &= s^2_{\alpha} + s^2_{\delta} \\
        &\sim \chi^2_{2(\rm{N_{src}}-1)}.
    \end{aligned}
\end{equation}

Using the data from 9 FRBs (see Table \ref{tab:frbs}), the above procedure was performed on each FRB. Individual $s^2$ values were calculated, fitting mean $\Delta[\alpha,\delta]$ values. The results of all $\chi^2$ tests are given in Table \ref{tab:frbs}. Only FRB~200430 and FRB~200906 have a sufficient number of sources to allow for a sensitive test of $\rm H_0$, and these reject the null hypothesis at high significance.

To perform a more accurate test, we also sum over the $s^2$ values of all FRBs to obtain the total variance
\begin{equation} \label{eq:tot_var}
\begin{aligned}
    s^2_{\rm tot} &= \sum_{j=1}^{\rm{N_{FRB}}} s^2_j  \\ 
    &\sim \chi^2_{\rm{NDF_{tot}}}
\end{aligned}
\end{equation}
\noindent where the total number of degrees of freedom (NDF) is defined as
\begin{equation}
    \rm{NDF_{tot}} \equiv \sum_{\textit{j}=1}^{\rm{N_{FRB}}} 2(\rm{N}_{\rm{src},\textit{j}} - 1), \label{eq:ndf_tot_def}
\end{equation}
\noindent and $j$ is the $j^{th}$ FRB.
This yields a one-sided p-value of $2.19 \times 10^{-12}$; that is, the estimated errors are smaller than the true errors at 7.0\,$\sigma$ significance.

Our findings could be due to the presence of systematic effects in the data. Potential systematics are likely to arise from effects such as unmodelled mismatched source structure (due to differences in the frequency and angular resolution in the compared observations), which would act to shift the fitted CRAFT field source centroid relative to that of the reference source; or directional dependencies in the offsets (e.g., due to a wedge in the ionosphere). Both would result in the additional error ($d[\alpha,\delta]^e$) not being centred at zero. Thus, we would not expect $\rm H_0$ to accurately model real-world observations.

We therefore discard $H_0$ and examine two further reasonable alternatives. Ordering by complexity, the hypotheses are:
\begin{itemize}
    \item $\rm H_1$: All uncertainties are equal but unknown; i.e., $\sigma^t = C$. In the limit in which the measurement uncertainty becomes negligible compared to unmodelled systematic effects (i.e., all field sources have a high S/N), we would expect $\rm H_1$ to be satisfied, as we would reach a `floor' in the attainable precision set by systematic contributions unrelated to the measurement S/N. The simplest possible form of such a systematic error floor would be a constant independent of the source.
    \item $\rm H_2$: The true uncertainties are proportional to the estimated uncertainties, with a constant of proportionality to be estimated from the data; i.e., $\sigma^t_i = C \sigma^e_i$. This provides the simplest possible way to include the effects of unmodelled  error contributions to the field source offsets and avoid the underestimation of the total uncertainty (on average) that would result from neglecting them.
\end{itemize}
\noindent In the following, we outline the results of testing each in turn.

Assuming $\rm H_1$ to be true, we first calculate unweighted mean offsets. Each resulting residual offset $r$ is calculated as
\begin{equation} \label{eq:residuals}
    r_{[\alpha,\delta]} = [\alpha,\delta]_{i,j}^e - \Delta[\alpha,\delta]_j,
\end{equation}
\noindent and then scaled to account for the residual-minimising effect of the procedure employed (i.e., the reduction in error due to the points being used to estimate the mean) when fitting the residuals versus estimated uncertainty
\begin{equation}
    r^{\prime}_{[\alpha,\delta]} = r_{[\alpha,\delta]} \sqrt{\frac{\rm{N_{src}}}{\rm{N_{src}}-1}}. \label{eq:ndf_reduction}
\end{equation}
We then determine if these scaled residuals show any dependence on the estimated uncertainties, which would reject $\rm H_1$ (i.e., the assumption that the distribution of offsets is a constant). A complexity arises, however, because three or more sources per FRB are required to obtain any meaningful data. Two sources, for instance, will always result in the fitted mean being halfway between them, and so the result will be independent of the true uncertainty. Thus, only data from FRBs with $\rm{N_{src}} > 2$ are included in the fit. 

The scaled residuals versus estimated uncertainty ($\sigma^e$) results were fit using standard linear regression (using the \textsc{linregress} function in \textsc{SciPy}) using all data as well as RA and Dec.\ independently. These fitted lines are also unweighted since the scatter should be independent of $\sigma^e$ under the assumption of $\rm H_1$. We find
\begin{equation}
    \begin{aligned}
        r^{\prime}_{\alpha,\delta} &= 1.76\sigma^e_{\alpha,\delta} - 0.3 \\
        r^{\prime}_{\alpha} &= 0.88\sigma^e_{\alpha} + 0.2 \\
        r^{\prime}_{\delta} &= 2.34\sigma^e_{\delta} - 0.7.
    \end{aligned}
\end{equation}
Under $\rm H_1$, the true slope is zero. Testing for consistency with this hypothesis, the (two-sided) p-values for our fitted slopes are then $3.5 \times 10^{-3}$ and $2.6 \times 10^{-6}$ for the RA- and Dec.-only fittings, respectively. Simultaneously fitting RA and Dec.\ yields a p-value of $1.0 \times 10^{-8}$. Therefore, there is very strong information that $\sigma^t$ and $\sigma^e$ are positively correlated, which is as expected. Therefore, we discard $\rm H_1$ and proceed to examine $\rm H_2$.

According to $\rm H_2$, we calculate the weighted mean offsets in RA and Dec.\ using Equation \ref{eq:estimates}. It can be shown that for a weighted mean, $\Bar{x}$, the expected deviation of the $i^{th}$ data point $x_i$ from that mean is given by
\begin{eqnarray}
\left< (x_i - \bar{x})^2 \right> & = & \frac{1}{w_i} - \frac{1}{\sum_k^{N_{\rm src}} w_k} \nonumber \\
& = & \sigma_i^2 - \frac{1}{\sum_k^{N_{\rm src}} \frac{1}{\sigma_k^2}}.
\end{eqnarray}
That is, the expected variance between a point and its estimated mean is not reduced by the usual $(\rm{N_{src}}-1)/\rm{N_{src}}$ factor of Equation~\ref{eq:ndf_reduction} but rather a factor $\eta$:
\begin{equation}
    \eta_i = 1 - \frac{1}{\sigma_i^2 \sum_k^{N_{\rm src}} \frac{1}{ \sigma_k^2}}.
\end{equation}
The residuals defined according to Equation \ref{eq:residuals} must therefore be multiplied by $\eta^{-0.5}$:
\begin{equation}
   r^\prime_{[\alpha,\delta]} = \frac{1}{\sqrt{\eta}} r_{[\alpha,\delta]} \label{eq:wrprime} 
\end{equation}
\noindent in order for them to be unbiased estimates of their standard deviation. This results in a factor of between 1 and 2 in this work.

We use linear regression to obtain weighted fits of the residuals $r^{\prime}$ as a function of their estimated uncertainty and find
\begin{equation}
    \begin{aligned}
        r^{\prime}_{\alpha,\delta} &= 1.43\sigma^e_{\delta} \\
        r^{\prime}_{\alpha} &= 1.40\sigma^e_{\delta} \\
        r^{\prime}_{\delta} &= 1.67\sigma^e_{\delta}.
    \end{aligned}
\end{equation}

The best-fit value of $C$ is 1.43. If $\sigma^e$ is correct, then the best-fit value of the slope of a fit to the residuals is expected to be:
\begin{eqnarray}
<C> & = & \frac{2 \int_0^{\inf} r p(r) dr}{2 \int_0^{\inf} p(r) dr}, \label{eq:expected_r} \\
p(r) & = & \frac{1}{\sigma^t \sqrt{2 \pi}} \exp\left\{-\frac{r^2}{{2(\sigma^t})^2}\right\}.
\end{eqnarray}
\noindent Here, $p(r)$ is the assumed Gaussian distribution of the errors. However, since the fits are against $|r|$, the integrals in Equation \ref{eq:expected_r} are evaluated from 0 to infinity. The denominator evaluates to unity since it is a normalised probability distribution. The numerator of Equation \ref{eq:expected_r} is
\begin{eqnarray}
&2& \int_0^{\inf} r \frac{1}{\sigma^t \sqrt{2 \pi}} \exp \left\{-\frac{r^2}{{2(\sigma^t})^2}\right\} dr \nonumber \\
&=& \sigma^t \sqrt{\frac{2}{\pi}}.
\end{eqnarray}
\noindent In other words, while under $\rm H_2$, using the weighted mean yields the correct best-fit position, but we expect the mean of the errors to underestimate $\sigma^t$ by a factor of $\sqrt{\frac{2}{\pi}}\approx 0.80 $.

It therefore appears that treating $\sigma^t = C \sqrt{\pi/2} \sigma^e \approx 1.25 C \sigma^e$ is correct. Given we find a best-fit of $C=1.43$, the best-fit true uncertainty is $\sigma^t = 1.79 \sigma^e$.

While our current dataset fails to reject $\rm H_2$ and found reasonable consistency with its predictions, this model can result in either over- or under-estimating the true uncertainties, depending on the characteristics of the continuum source sample in a given field. Additionally, we can only confirm cases in which over- or under-estimation has occurred in fields with sufficient background sources; with only a handful or sources, it is impossible to determine if this has taken place and to what degree.

Thus, while $\rm H_2$ is a reasonable alternative to the currently used simple weighted mean method of estimating the mean image frame offsets and uncertainties and is testable given the current sample size of localised FRBs, the model should be further refined as the number of usable FRBs grows. One such refinement would effectively combine $\rm H_1$ and $\rm H_2$ -- i.e., introducing a systematic term ($A$) in addition to the random estimated uncertainty: $\sigma^t_i=\sqrt{A^2+C^2 (\sigma^e_i)^2}$. This might more accurately capture the systematic effects known to potentially exist in the data (e.g., source structure and directional dependencies in the individual offsets). In addition, future models could be parameterised over a given range appropriate to a particular catalogue or instrument, thereby accounting for the characteristics of the reference used for the comparison (e.g., the astrometric accuracy limits of a catalogue). In addition, the astrometric measurements outlined in this work could be performed on random fields with several sources (rather than on single sources), which would increase the number of degrees of freedom and facilitate multiple individual tests of a given model. This would, for example, enable consistency crosschecks of the scale factor derived for $\rm H_2$.

Until this becomes possible, we take $\rm H_2$ as our working hypothesis, namely that the best estimate of the offset is given by a weighted mean according to the estimated errors $\sigma^e$ but that the magnitude of these errors -- and consequently the estimated error in the weighted mean -- should be scaled up by a factor of 1.79.

\subsubsection{Deriving Updated Positional Information} \label{sec:updated_frbs}

Table \ref{tab:posoffunc} lists the published positions, offsets, and uncertainties for each FRB in our sample along with the updated uncertainties obtained when using the above derived scale factor. While the scale factor only changes the systematic uncertainty estimation, we have also updated the positions and offsets for the published FRBs that were affected by the formerly incorrect weighting scheme detailed in \cite{Macquart2020_DMz}, which used $w_i = 1/\sigma^e_{[\alpha,\delta],i}$ as the weights in Equation \ref{eq:estimates} rather than these values squared.

This change only affects FRBs that have both more than one field source used for the comparison (i.e., those for which this weighted mean method can be used) and offsets that were not originally consistent with zero. Therefore, the positions and offsets for FRBs~180924, 181112, and 190102 are unchanged, while the uncertainties are scaled by the factor derived in Section \ref{sec:field_src_offset_deriv}. Conversely, the positions, offsets, and uncertainties for FRBs~190608 (no change in position to the quoted precision), 190611, 190711, and 190714 have been updated using both the correct weights (affecting all positional information) and the scale factor (further modifying the uncertainties). The FRB~191001 position has been updated both to account for this new weighting scheme and to rectify an error in the original RA offset correction reported in \cite{Bhandari2020ApJ...901L..20B}.

FRB~200430 \citep{Heintz_hostII_2020ApJ...903..152H} includes both the offsets and uncertainties derived via the nominal method along with an additional offset and uncertainty in Dec.\ to account for frequency dependence in the position (Section \ref{sec:200430}), and so the nominal components of the offsets and uncertainties are updated, including the use of the scale factor derived in Section \ref{sec:field_src_offset_deriv}), resulting in an overall change in the position and the estimated offsets and uncertainties. Since the positional information for both FRB~191228 and FRB~200906 (Bhandari et al. in prep) was obtained by using the updated weights and the scale factor, this information is unchanged.

We also provide the revised total astrometric uncertainties (i.e., the quadrature sum of the statistical and systematic uncertainties) in RA and Dec.\ for each FRB in our sample.

\begin{table*}
\caption{Published FRB positions ($[\alpha,\delta]_{\rm{pub}}$), weighted mean offsets ($\Delta[\alpha,\delta]_{\rm{pub}}$), and systematic uncertainties ($\sigma_{[\alpha,\delta]_{\rm{pub}}}$) and their revised values (where updates are required and denoted by the subscript rev) as per the work in Section \ref{sec:field_src_offset_deriv}. Finally, we list the total revised uncertainty in RA and Dec., $\sigma_{[\alpha,\delta],\rm{tot}}$ (i.e., the quadrature sum of the statistical [not shown] and updated systematic uncertainties). We note that the precision of the uncertainties is given such that it matches that reported in the references listed in Table \ref{tab:frbs}, with the RA precision including an additional significant figure to mitigate round-off errors when converting to seconds.}
\centering
\begin{threeparttable}
\begin{tabular}{lrrrrrrr}
\hline\hline
\multicolumn{1}{l}{Source} & \multicolumn{1}{c}{$[\alpha,\delta]_{\rm{pub}}$\tnote{$\dagger$}} & \multicolumn{1}{c}{$[\alpha,\delta]_{\rm{rev}}$} & \multicolumn{1}{c}{$\Delta[\alpha,\delta]_{\rm{pub}}$} & \multicolumn{1}{c}{$\Delta[\alpha,\delta]_{\rm{up}}$} & \multicolumn{1}{c}{$\sigma_{[\alpha,\delta]_{\rm{pub}}}$} & \multicolumn{1}{c}{$\sigma_{[\alpha,\delta]_{\rm{rev}}}$\tnote{$\ddag$}} & \multicolumn{1}{c}{$\sigma_{[\alpha,\delta],\rm{tot}}$} \\
 & & & (arcsec) & (arcsec) & (arcsec) & (arcsec) & (arcsec) \\
\hline%
  FRB~180924 & 21h44m25.255s & & 0.0 & & 0.0900 & 0.1611 & 0.1756 \\
   &  $-$40d54m00.10d & & 0.0 & & 0.09 & 0.16 & 0.18 \\
  FRB~181112 & 21h49m23.63s & & 0.0 & & 2.150 & 3.849 & 3.875 \\
   &  $-$52d58m15.4s & & 0.0 & & 1.4 & 2.4 & 2.4 \\
  FRB~190102 & 21h29m39.76s & & 0.0 & & 0.440 & 0.788 & 0.805 \\
   & $-$79d28m32.5s & & 0.0 & & 0.5 & 0.9 & 1.0 \\
  FRB~190608 & 22h16m4.77s & 22h16m04.77s & 0.41 & 0.46 & 0.185 & 0.327 & 0.380 \\
   & $-$07d53m53.7s & $-$07d53m53.7s & $-$0.90 & $-$0.89 & 0.2 & 0.3 & 0.4 \\
  FRB~190611 & 21h22m58.91s & 21h22m58.94s & 1.67 & 1.74 & 0.629 & 1.119 & 1.164 \\
   & $-$79d23m51.3s & $-$79d23m51.3s & 0.25 & 0.24 & 0.6 & 1.1 & 1.1 \\
  FRB~190711 & 21h57m40.68s & 21h57m40.62s & 1.7 & 1.5 & 0.381 & 0.646 & 0.657 \\
   & $-$80d21m28.8s & $-$80d21m28.8s & $-0.4$ & $-$0.4 & 0.3 & 0.6 & 0.6 \\
  FRB~190714 & 12h15m55.12s & 12h15m55.13s & 0.71 & 0.92 & 0.32 & 0.52 & 0.54 \\
   & $-$13d01m15.7s & $-$13d01m15.6s & $-$1.45 & $-$1.35 & 0.23 & 0.38 & 0.4 \\
  FRB~191001 & 21h33m24.373s & 21h33m24.313s & 0.731 & 0.765 & 0.1073 & 0.1737 & 0.2144 \\
   & $-$54d44m51.86s &  $-$54d44m51.86s & $-$0.808 & $-$0.811 & 0.10 & 0.16 & 0.18 \\
  FRB~191228\tnote{$*$} & 22h57m43.24s & & 0.410 & & 0.830 & & 0.899 \\
   & $-$29d35m37.0s & & $-$0.856 & & 0.823 & & 0.890 \\
  FRB~200430 & 15h18m49.54s & 15h18m49.54s & $-$0.03 & $-$0.04 & 0.2500 & 0.2506 & 0.3015 \\
   & 12d22m36.8s & 12d22m36.3s & 4.12 & 3.62 & 1.04 & 0.98 & 1.01 \\
  FRB~200906\tnote{$*$} & 03h33m59.08s & & 2.05 & & 0.34 & & 0.35 \\
   & $-$14d04m59.5s & & 0.51 & & 0.55 & & 0.56 \\
\hline\hline
\end{tabular}
\begin{tablenotes}[flushleft]
        \item[$\dagger$] Published positions are from the same references listed in Table \ref{tab:frbs}.
        \item[$\ddag$] For offsets consistent with zero, the updated uncertainties are simply $1.79 \times \sigma_{[\alpha,\delta]_{\rm{pub}}}$. Otherwise, they have been re-derived, if applicable, with both the updated weighting scheme in Equation \ref{eq:estimates} and the use of the scale factor.
        \item[$*$] The values for FRB~191228 and FRB~200906 (Bhandari et al. in prep) were derived using the method described in Section \ref{sec:field_src_offset_deriv}. For these reasons, the updated value columns for these FRBs are intentionally left blank.
    \end{tablenotes}
\end{threeparttable}
\label{tab:posoffunc}
\end{table*}

\subsection{FRB~200430: The Case of Frequency-Dependent Offsets} \label{sec:200430}

FRB~200430, which was detected at a central frequency of 863.5~MHz, shows a frequency dependence in its offsets, which has not been seen in any other FRBs to date. \cite{Heintz_hostII_2020ApJ...903..152H} reported the detection and briefly outlined the steps taken to account for the bias introduced by this frequency dependence when using the field data to estimate the offsets and uncertainties in RA and Dec.\ for this burst. We expand this description here and compare the data presented in this work to that of FRB~200430.

In determining the final statistical position (i.e., prior to any offset correction) for a given FRB, an optimal slice of the data roughly centred on the temporal position of the FRB (the `gated' data) is correlated, and the subsequently calibrated visibilities are then re-weighted by a spectrum derived from the cube-imaged data (using pixels covering the peak FRB emission) to boost the S/N across the band \citep[see, e.g., ][for a full description]{Bannister565}. In doing so for FRB~200430, we found both a shift and larger statistical uncertainty in the fitted Dec.\ $>$1-$\sigma$, both of which are indicative of phase/systematic errors in the Dec.\ as a function of frequency. In addition, we also measured significant offsets in Dec.\ (but not RA) in the background field sources relative to their FIRST counterparts, which likewise signals the presence of phase errors in the calibration data. We note that the same re-weighting is not used for the field data, and so, since the systematic offsets in the re-weighted FRB image frame cannot be corrected for with the field sources, this precluded us from using the re-weighted FRB position.

In order to confirm the presence of the suspected frequency dependence in Dec., we made a cube image of the optimally gated FRB data with a resolution of 56~MHz (i.e., 1/6th of the 336-MHz bandwidth), which appeared to show a drift in Dec.\ as a function of frequency. The FRB position in each of the 6 channels of the image was fitted via JMFIT in the manner described in Section \ref{sec:software} \citep[see also, e.g., ][]{Day2020}. We note that, since this FRB is brighter at lower frequencies, the statistical uncertainties on these positions also increase with frequency. While there was a slight, non-frequency-dependent offset in RA ($\lesssim 1\sigma$ and roughly accounted for by the increasing uncertainties), there was an $\sim~7$-arcsec offset in Dec.\ across the band (i.e., the the positions at the band edges were inconsistent at the 2.5-$\sigma$ level), with a clear dependence on frequency that cannot be accounted for with the increased uncertainties.

After confirming that the calibration data visibilities used for the FRB data showed the expected nominal properties (i.e., amplitudes $\sim$ 15 Jy [J1939] and phases centred around zero), the final calibrator scan ($\sim$ 20 minutes after the first) was correlated to determine if it showed the same systematic offsets as a function of frequency. In performing the same steps as before, we found a similar non-frequency-dependent offset in RA and a slightly worse, frequency-dependent offset in Dec.\ (i.e., the the positions at the band edges were inconsistent at the 3-$\sigma$ level). Notably, when comparing the positions derived using the two scans to calibrate the FRB, the overall differential offset in each channel was stable as a function of frequency.

We also investigated if any differential phase offsets were present across the 20-minute time span between the two calibrator scans, where phase changes $\gtrsim 1$ degree would give rise to non-negligible calibration errors. We calibrated the first calibrator scan with the last and then ran a \textsc{gaincal} in \textsc{casa} in the phase solution mode to determine a single phase offset (averaging the two polarisations) per antenna. Phase offsets of $\lesssim 10$~degrees were found, with a mean of approximately 2 degrees. We therefore concluded that there were phase errors in one or both of these calibrator scans. However, since the better scan cannot be conclusively determined, we attempted to correct for this frequency dependence using the first scan as detailed in the following.

The typical spectral index of a continuum field source is $\sim -0.7$ (i.e., that of synchrotron radiation), while the FRB~200430 spectral index appeared to be much steeper from the initial spectrum. Since the offset in the FRB position is a function of frequency and given the differing spectral indices in the field sources versus the FRB, the field source centroids would not be affected in the same way by the frequency-dependent phase errors, and thus our nominal method of correction using the field sources would introduce a bias.

In order to account for this, we derived a coarse spectral index for the FRB and compared this to the typical field source spectral index. To calculate the FRB spectral index ($\gamma_{\rm FRB}$), using standard linear regression, we used the coarse cube of the FRB to fit the log of the extracted flux densities ($S$) versus frequency ($\nu$), given by
\begin{equation}
    \mathrm{log} S = \mathrm{log} A + \gamma \mathrm{log} \nu,
\end{equation}
\noindent where $A$ is a constant of proportionality and $\gamma$ is the spectral index. We found $\gamma_{\rm FRB} \approx -5.46$.

The final images made from both the FRB and field datasets must be frequency averaged in order to maximise the S/N of source detections in all images and thereby the astrometric accuracy attainable. Given $S = A\nu^{\gamma}$, the central frequency ($\nu_{\rm cen}$) is defined as the frequency at which the area under the curve is 50\% of the total area from the lowest ($\nu_1$) to highest ($\nu_2$) frequencies, and this corresponds to the frequency at which the centroid of the averaged image positions will be located. These quantities are given by
\begin{equation} \label{eq:cen_freq_def}
\begin{aligned}
    \int_{\nu_1}^{\nu_{\mathrm{cen}}} \nu^{\gamma} &= 0.5 \int_{\nu_1}^{\nu_2} \nu^{\gamma}, \\
    \implies \nu_{\rm cen} &= \big[0.5\big(\nu_1^{\gamma+1} + \nu_2^{\gamma+1}\big)\big]^{1/(\gamma+1)}
\end{aligned}
\end{equation}

Due to the frequency dependent offset in Dec.\ and the differing spectral indices, the central frequencies in the two images differ. We therefore derived a typical deviation as a function of frequency, which could then serve as the uncertainty expected at a given frequency due to the introduced bias. We assumed a typical spectral index of $-0.7$ for the field sources and found $\nu_{\rm cen,field} \approx 855.46$~MHz. Using the derived value of $\gamma_{\rm FRB} = -5.5$, we found a central frequency for the FRB of $\nu_{\rm cen,FRB} \approx 806.38$~MHz. Thus, the frequency difference between the centroid locations in the two images is $\Delta\nu_{\rm cen} \sim 49$~MHz.

In order to determine the expected offset in the position for the offset in Dec., we used linear regression to obtain a weighted fit of the Dec.\ values measured in the 56-MHz resolution cube image versus frequency. We found offsets at the respective central frequencies of the FRB and field images of $\Delta\delta_{\rm cen,FRB} = 1.39$~arcsec and $\Delta\delta_{\rm cen,field} = 2.32$~arcsec. Thus, for a central frequency offset of $\sim$49~MHz, we found an offset of 0.93~arcsec (field to FRB). This was then added to the FRB Dec.\ position in addition to the offset derived via estimating the nominal offsets and uncertainties based on the scatter in the field source offsets. The originally reported position in \cite{Heintz_hostII_2020ApJ...903..152H} used nominal weighted mean values of $-0.03 \pm 0.25$~arcsec and $3.19 \pm 0.47$~arcsec for RA and Dec., respectively. However, as detailed in Section \ref{sec:updated_frbs}, the weights used to derive them were incorrect, and so we have updated these values using the corrected weighting scheme to $-0.04 \pm 0.25$~arcsec for RA and $2.69 \pm 0.30$~arcsec for Dec. We conservatively estimated the uncertainty on the offset due to the frequency dependence of the FRB position to be equivalent to the offset correction (i.e., $0.93 \pm 0.93$). Summing the two sources of systematic offsets in Dec.\ and combining their uncertainties in quadrature, we found a total systematic Dec.\ offset and uncertainty of $3.62 \pm 0.98$~arcsec. These offsets were used to correct the FRB position, yielding a final RA $=$ 15h18m49.54s $\pm$ 0.021 (statistical; systematic: $\pm$0.011 s; $\pm$0.017 s) and
Dec.\ $=$ 12d22m36.3s $\pm$ 1.01 (statistical; systematic: $\pm$0.24; $\pm$0.98). We note that the RA is unchanged from the previously published value (Table \ref{tab:posoffunc}) at the quoted precision. Given the significant impact on the final position of the frequency-dependent offset observed for this FRB, all future FRBs, especially those at low frequencies, should be inspected to determine if such offsets exist in the data.

While we have not seen this frequency dependence in the position of any other FRBs, as discussed in Section \ref{sec:res_analysis}, we do see a dependence on wavelength in the low-band data presented in this work. FRB~200430 was also detected in the lower frequency range observable with CRAFT, and while the gradient of this offset dependence is larger for this burst than that seen in Figure \ref{fig:beam30_offvs.time_freq_lowband}, it is not inconsistent with our data, in which we see offset changes of order a few arcsec across the band.

A possible contributor to the more extreme offset gradient exhibited by FRB~200430 is the ionosphere. We therefore investigated its likely contributions to the total systematic offset. The $\sim$7-arcsec shift in the Dec.\ across the band is about a quarter of the beam (i.e., $\sim 90$ degrees), which is approximately 0.15 total electron content units (TECU) of difference in the differential ionosphere across the array. \cite{Mevius_2016} report measurements from the LOw‐Frequency Radio interferometer ARray (LOFAR) of short timescale variations on ionospheric sightlines on of order km baselines. These show that we should not typically see variation $\sim 0.15$ TECU across the array on baselines out to 6~km. However, variations roughly 5x smaller (corresponding to offsets of order 1 arcsec across the band at these frequencies) do occur. While these LOFAR measurements were taken at a Dec.\ of $+50$, the ionosphere at declinations observable with ASKAP is not expected to be significantly different, and indeed, the mid band data we present in this work shows offsets due to the ionosphere of up to $\sim 1$~arcsec in RA and $\sim 0.4$~arcsec in Dec., resulting from increased ionospheric activity during these observations (Section \ref{sec:ionocal}). Likewise, if extrapolating the mid-band results, similar conditions during the low-band observations would have led to offsets up to $\sim 2$~arcsec in RA and $\sim1$ arcsec in Dec. Thus, the $\sim 7$~arcsec shift in the FRB~200430 position across the band is much larger than the expected ionospheric contribution based on the tests presented in this work and the LOFAR measurements \citep{Mevius_2016}. Moreover, if the ionosphere were the dominant component, the frequency dependence in the offsets would be better fit by a quadratic rather than a linear model. Thus, ionospheric effects cannot solely account for the observed offset across the band.

\section{Conclusion} \label{sec:conc}

We have presented a method for and the results of estimating the typical astrometric accuracy of positions obtained via the snapshot imaging technique. Using a set of strong compact sources observed with the CRAFT and ASKAP systems, we calculate offsets between the fitted and reference source positions. We find that the offset distributions we estimate match the published FRB offset distributions well in both bands (Figure \ref{fig:tru_dist}). We also note a weak dependence of the offsets on differential time and elevation (i.e., relative to the calibrator scan), with a great deal of scatter (Figures \ref{fig:beam30_offvs.tel_lowband} and \ref{fig:beam30_offvs.tel_highband}). These trends are clearer in Dec.\ than RA, and the scatter is generally more significant in the low-band. We detect no trend in the offsets with angular separation from the calibrator, but we note that future studies with an expanded sample would likely better quantify and account for these dependencies. We do, however, find a significant frequency dependence in the low-band data, which is consistent with linear growth with wavelength and therefore a frequency-independent phase error unmodelled by the calibration solutions. We also find that all these results are consistent across the sampled beams and when using both natural and Briggs weighting schemes for the imaging.

In modelling the large-scale effects of the ionosphere, we detected increased activity on the day the mid-band observations were conducted, resulting in higher differential offsets versus the low-band observations when comparing the positional offsets estimated when using the nominal versus the ionosphere-corrected calibration solutions. We conclude that the ionosphere might contribute to the temporal and spatial offsets we measure, but the extent of any contribution requires further study.

We also show the results of investigating various models of increasing complexity that can be used to estimate the systematic offset and its uncertainty in the frame registration as alternatives to the current method of using a simple weighted mean for the former and either the error in this mean or the scatter in the measured field source offsets about the mean for the latter. We find that an initial hypothesis assuming the true offset uncertainties are well estimated by the measured uncertainties (i.e., no systematics) is not well supported by the data. Since systematics such as source structure and directional dependence in the offsets are known to occur, this model rejection is expected. We also test a model in which all uncertainties are equal to some constant but unknown; that is, the true uncertainties are independent of the measured uncertainties. We find, however, that there is strong evidence that the uncertainties are dependent on the estimated uncertainties, as we would expect, and so we likewise reject this model. Finally, we model the true uncertainties as proportional to the measured uncertainties and find good agreement with the data and the model predictions. We note that while this model can result in both over- and under-estimation of the uncertainties, depending on the sources within the sample for any given field, it is a reasonable alternative to the current method and testable given our current sample of localised FRBs. Of note, future studies with a larger sample size, including dedicated observations of fields with multiple sources (see Section \ref{sec:field_src_offset_deriv}), would facilitate testing more complex models, thereby enabling long-term improvements to our estimations of the systematic uncertainty.

When comparing the positional offsets derived using the ASKAP hardware correlator data and the CRAFT software correlator data, we find that the former track the latter very well (Figures \ref{fig:beam30_offvs.tel_lowband} and \ref{fig:beam30_offvs.tel_highband}). Given the higher S/N, due to the longer integration time, in the images made from the ASKAP hardware correlator data and the relationship between S/N and astrometric accuracy (i.e., improving the former improves our estimation of the latter), the use of the hardware correlator data to derive calibration solutions and perform the frame registration for commensal observations, in which the hardware correlator data are available, is a promising future avenue we have investigated here. In particular, the calibration solutions derived for the hardware correlator data could be applied to the software correlator data, and the higher S/N field sources detected in the image made from the hardware correlator data could then be compared to their counterpart reference positions to obtain any residual frame offsets in RA and Dec.\ between the ASKAP frame and the ICRF3.

In comparing the CRAFT- and ASKAP-derived offset distributions, we have shown that the typical residual systematic offsets between these image frames (i.e., the typical error expected when applying the hardware correlator data calibration solutions to the software correlator data) fall in the nominal ranges of $\sim 0.5 - 0.6$~arcsec (low-band) and $\sim 0.2 - 0.3$~arcsec (mid-band), for the naturally weighted image case and when combining the central 68\% of the samples from the RA and Dec.\ offset distributions in the simple case of performing only a bandpass and phase calibration (Figure \ref{fig:hw_applied}). We find that when using Briggs weighting with a robustness of 0.0, these residuals improve by up to 17\% in the low band and 33\% in the high band [in the high-S/N, reasonable ($u,v$) coverage, and wide elevation range study we have conducted here]. Thus, when applying the hardware correlator-derived calibration solutions to the FRB image and using the higher S/N hardware correlator data to perform the frame registration, Briggs weighting is preferred. Furthermore, although the residual offsets between the two image frames are not constant, there is no trend in these residuals with time, elevation, or angular separation, and so we find that these residual frame offsets could be reasonably applied to the CRAFT image frame as an additional component of the systematic astrometric uncertainty when ASKAP calibration solutions are transferred to the CRAFT data regardless of the time, elevation, or angular separation between the target (FRB) and calibrator.

In the case of non-negligible offsets between the ASKAP and reference frames (i.e., due to the presently known astrometric offsets in the ASKAP data), the estimated systematic offset uncertainty between these frames would then be combined with both the residual CRAFT-ASKAP frame offset uncertainty and the statistical positional uncertainty of the FRB to estimate the total positional uncertainty. (We note that the statistical uncertainty in the ASKAP field source positions is expected to be much smaller than the CRAFT-ASKAP residual uncertainty in the limit of high S/N, which is always the case with the hardware data.) Conversely, if the error in registering the ASKAP frame to the ICRF3 can be sufficiently reduced such that these offsets become negligible (i.e., when ASKAP is shown to be well registered), the residual CRAFT-ASKAP frame offsets would then dominate the estimated systematic uncertainty. In this case, the residual CRAFT-ASKAP offset is then a reasonable estimate of the typical total systematic uncertainty, which would then be combined as usual with the statistical uncertainty to obtain the final astrometric uncertainty in the FRB position. Since this is reasonably well-watched to the typical statistical uncertainty in the FRB positions, without the need to also correlate the CRAFT data to image the field, we can use this approach to do sub-galaxy localisations out to a moderate redshift.

Finally, we explore the case of frequency-dependent Dec.\ offsets in the low-band-detected FRB~200430 \citep{Heintz_hostII_2020ApJ...903..152H} and a method of estimating the additional systematic offset and uncertainty introduced by using the snapshot technique. We measured an offset in the FRB Dec.\ of approximately 7 arcsec across the 336-MHz band. In addition to this, the Dec.\ (but not RA) offsets found in the field source positions and the significant differential phase offset between calibrator scans separated by 20 minutes all indicated the presence of residual phase errors in the calibration solutions. Using a 56-MHz resolution cube image of the FRB, we estimated a spectral index, compared this to the typical spectral index of a continuum field source, and determined a typical Dec.\ offset due to the central frequency offset in the frequency-averaged images to account for the bias introduced when using the field sources to correct the FRB image frame. We then take this as both the estimated offset and uncertainty due to this bias and combine these with the values derived from the nominal frame registration method. While measurements of the differential ionosphere with LOFAR predict offsets of order 1 arcsec across the band (at $\sim 864$-MHz) and on $\sim$6-km baselines \citep{Mevius_2016}, this does not account for the bulk of the shift we detect. However, while this is an extreme case, the magnitude of these offsets is not unreasonable given the frequency dependence detected in the data presented in this work. Given both of these results, all future FRB data used for localisations should be checked for any frequency dependence, especially when detected at low frequencies where such effects can heavily influence the accuracy of the position.

\begin{acknowledgements}
The Australian SKA Pathfinder is part of the Australia Telescope National Facility which is managed by the Commonwealth Scientific and Industrial Research Organisation (CSIRO). Operation of ASKAP is funded by the Australian Government with support from the National Collaborative Research Infrastructure Strategy. ASKAP uses the resources of the Pawsey Supercomputing Centre. Establishment of ASKAP, the Murchison Radio-astronomy Observatory and the Pawsey Supercomputing Centre are initiatives of the Australian Government, with support from the Government of Western Australia and the Science and Industry Endowment Fund. We acknowledge the Wajarri Yamatji people as the traditional owners of the Observatory site.
A.T.D. is the recipient of an Australian Research Council Future Fellowship (FT150100415).
RMS acknowledges support through Australian Research Council Future Fellowship FT190100155.
K.B. acknowledges Australian Research Council (ARC) grant DP180100857.
We also thank Maxim Voronkov and Evan Keane for their helpful feedback.
\end{acknowledgements}

\bibliographystyle{pasa-mnras}
\bibliography{ast}

\begin{thebibliography}{}
\makeatletter
\relax
\def\mn@urlcharsother{\let\do\@makeother \do\$\do\&\do\#\do\^\do\_\do\%\do\~}
\definecolor{darkblue}{rgb}{0,0,0.597656}
\def\mndoi{\begingroup\mn@urlcharsother \@ifnextchar [ {\mndoi@} {\mndoi@[]}}
\def\mndoi@[#1]#2{\def\@tempa{#1}\ifx\@tempa\@empty \href
  {http://dx.doi.org/#2} {\textcolor{darkblue}{doi:#2}}\else \href
  {http://dx.doi.org/#2} {\textcolor{darkblue}{#1}}\fi \endgroup}
\def\mn@eprint#1#2{\mn@eprint@#1:#2::\@nil}
\def\mn@eprint@arXiv#1{\href {http://arxiv.org/abs/#1} {{\tt arXiv:#1}}}
\def\mn@eprint@dblp#1{\href {http://dblp.uni-trier.de/rec/bibtex/#1.xml}
  {dblp:#1}}
\def\mn@eprint@#1:#2:#3:#4\@nil{\def\@tempa {#1}\def\@tempb {#2}\def\@tempc
  {#3}\ifx \@tempc \@empty \let \@tempc \@tempb \let \@tempb \@tempa \fi \ifx
  \@tempb \@empty \def\@tempb {arXiv}\fi \@ifundefined
  {mn@eprint@\@tempb}{\@tempb:\@tempc}{\expandafter \expandafter \csname
  mn@eprint@\@tempb\endcsname \expandafter{\@tempc}}}

\bibitem[\protect\citeauthoryear{{Astropy Collaboration} et~al.,}{{Astropy
  Collaboration} et~al.}{2018}]{astropy:2018}
{Astropy Collaboration} et~al., 2018, \mndoi [\aj] {10.3847/1538-3881/aabc4f},
  \href {https://ui.adsabs.harvard.edu/abs/2018AJ....156..123A} {156, 123}

\bibitem[\protect\citeauthoryear{Bannister et~al.,}{Bannister
  et~al.}{2019}]{Bannister565}
Bannister K.~W.,  et~al., 2019, \mndoi [Science] {10.1126/science.aaw5903},
  365, 565

\bibitem[\protect\citeauthoryear{{Becker}, {White}  \& {Helfand}}{{Becker}
  et~al.}{1995}]{Becker_FIRSTsurvey1995ApJ...450..559B}
{Becker} R.~H.,  {White} R.~L.,   {Helfand} D.~J.,  1995, \mndoi [\apj]
  {10.1086/176166}, \href
  {https://ui.adsabs.harvard.edu/abs/1995ApJ...450..559B} {450, 559}

\bibitem[\protect\citeauthoryear{{Bhandari} et~al.,}{{Bhandari}
  et~al.}{2020a}]{Bhandari2020_hostgalaxies}
{Bhandari} S.,  et~al., 2020a, \mndoi [\apjl] {10.3847/2041-8213/ab672e}, \href
  {https://ui.adsabs.harvard.edu/abs/2020ApJ...895L..37B} {895, L37}

\bibitem[\protect\citeauthoryear{{Bhandari} et~al.,}{{Bhandari}
  et~al.}{2020b}]{Bhandari2020ApJ...901L..20B}
{Bhandari} S.,  et~al., 2020b, \mndoi [\apjl] {10.3847/2041-8213/abb462}, \href
  {https://ui.adsabs.harvard.edu/abs/2020ApJ...901L..20B} {901, L20}

\bibitem[\protect\citeauthoryear{Briggs}{Briggs}{1995}]{briggs_thesis}
Briggs D.~S.,  1995, {PhD} dissertation, New Mexico Inst. of Mining \&
  Technology

\bibitem[\protect\citeauthoryear{Chatterjee et~al.,}{Chatterjee
  et~al.}{2017}]{Chatterjee17}
Chatterjee S.,  et~al., 2017, \mndoi [Nature] {10.1038/nature20797}, 541, 58

\bibitem[\protect\citeauthoryear{{Day} et~al.,}{{Day} et~al.}{2020}]{Day2020}
{Day} C.~K.,  et~al., 2020, \mndoi [\mnras] {10.1093/mnras/staa2138}, \href
  {https://ui.adsabs.harvard.edu/abs/2020MNRAS.497.3335D} {497, 3335}

\bibitem[\protect\citeauthoryear{{Day}, {Bhandari}, {Deller}, {Shannon}  \&
  {ASKAP-CRAFT Survey Science Project}}{{Day}
  et~al.}{2021}]{20201124a_low-band_ATel2021ATel14592....1D}
{Day} C.~K.,  {Bhandari} S.,  {Deller} A.~T.,  {Shannon} R.~M.,   {ASKAP-CRAFT
  Survey Science Project} 2021, The Astronomer's Telegram, \href
  {https://ui.adsabs.harvard.edu/abs/2021ATel14592....1D} {14592, 1}

\bibitem[\protect\citeauthoryear{Deller et~al.,}{Deller
  et~al.}{2011}]{Deller11}
Deller A.~T.,  et~al., 2011, \mndoi [PASP] {10.1086/658907}, 123, 275

\bibitem[\protect\citeauthoryear{{Fonseca} et~al.,}{{Fonseca}
  et~al.}{2020}]{CHIME9repeaters2020ApJ...891L...6F}
{Fonseca} E.,  et~al., 2020, \mndoi [\apjl] {10.3847/2041-8213/ab7208}, \href
  {https://ui.adsabs.harvard.edu/abs/2020ApJ...891L...6F} {891, L6}

\bibitem[\protect\citeauthoryear{{Gordon}}{{Gordon}}{2018}]{ICRF3_2018AGUFM.G42A..01G}
{Gordon} D.,  2018, in AGU Fall Meeting Abstracts. pp G42A--01

\bibitem[\protect\citeauthoryear{{Greisen}}{{Greisen}}{2003}]{Greisen03}
{Greisen} E.~W.,  2003, in {Heck} A.,  ed.,  Astrophysics and Space Science
  Library Vol. 285, Information Handling in Astronomy - Historical Vistas.
  p.~109, \mndoi{10.1007/0-306-48080-8_7}

\bibitem[\protect\citeauthoryear{{Heintz} et~al.,}{{Heintz}
  et~al.}{2020}]{Heintz_hostII_2020ApJ...903..152H}
{Heintz} K.~E.,  et~al., 2020, \mndoi [\apj] {10.3847/1538-4357/abb6fb}, \href
  {https://ui.adsabs.harvard.edu/abs/2020ApJ...903..152H} {903, 152}

\bibitem[\protect\citeauthoryear{{Hotan} et~al.,}{{Hotan}
  et~al.}{2014}]{Hotan2014PASA...31...41H}
{Hotan} A.~W.,  et~al., 2014, \mndoi [\pasa] {10.1017/pasa.2014.36}, \href
  {https://ui.adsabs.harvard.edu/abs/2014PASA...31...41H} {31, e041}

\bibitem[\protect\citeauthoryear{{Hotan} et~al.,}{{Hotan}
  et~al.}{2021}]{Hotan2021PASA...38....9H}
{Hotan} A.~W.,  et~al., 2021, \mndoi [\pasa] {10.1017/pasa.2021.1}, \href
  {https://ui.adsabs.harvard.edu/abs/2021PASA...38....9H} {38, e009}

\bibitem[\protect\citeauthoryear{{Law} et~al.,}{{Law}
  et~al.}{2020}]{Law2020ApJ...899..161L}
{Law} C.~J.,  et~al., 2020, \mndoi [\apj] {10.3847/1538-4357/aba4ac}, \href
  {https://ui.adsabs.harvard.edu/abs/2020ApJ...899..161L} {899, 161}

\bibitem[\protect\citeauthoryear{Macquart et~al.,}{Macquart
  et~al.}{2010}]{CRAFT_2010}
Macquart J.-P.,  et~al., 2010, \mndoi [PASA] {10.1071/AS09082}, 27, 272

\bibitem[\protect\citeauthoryear{{Macquart} et~al.,}{{Macquart}
  et~al.}{2020}]{Macquart2020_DMz}
{Macquart} J.~P.,  et~al., 2020, \mndoi [\nat] {10.1038/s41586-020-2300-2},
  \href {https://ui.adsabs.harvard.edu/abs/2020Natur.581..391M} {581, 391}

\bibitem[\protect\citeauthoryear{{Mannings} et~al.,}{{Mannings}
  et~al.}{2020}]{Mannings2020}
{Mannings} A.~G.,  et~al., 2020, ApJ, in press, \href
  {https://ui.adsabs.harvard.edu/abs/2020arXiv201211617M} {p. arXiv:2012.11617}

\bibitem[\protect\citeauthoryear{{Marcote} et~al.,}{{Marcote}
  et~al.}{2020}]{Marcote2020Natur.577..190M}
{Marcote} B.,  et~al., 2020, \mndoi [\nat] {10.1038/s41586-019-1866-z}, \href
  {https://ui.adsabs.harvard.edu/abs/2020Natur.577..190M} {577, 190}

\bibitem[\protect\citeauthoryear{{McConnell} et~al.,}{{McConnell}
  et~al.}{2020}]{RACS_I_2020PASA...37...48M}
{McConnell} D.,  et~al., 2020, \mndoi [\pasa] {10.1017/pasa.2020.41}, \href
  {https://ui.adsabs.harvard.edu/abs/2020PASA...37...48M} {37, e048}

\bibitem[\protect\citeauthoryear{{McMullin}, {Waters}, {Schiebel}, {Young}  \&
  {Golap}}{{McMullin} et~al.}{2007}]{CASA2007ASPC..376..127M}
{McMullin} J.~P.,  {Waters} B.,  {Schiebel} D.,  {Young} W.,   {Golap} K.,
  2007, in {Shaw} R.~A.,  {Hill} F.,   {Bell} D.~J.,  eds,  Astronomical
  Society of the Pacific Conference Series Vol. 376, Astronomical Data Analysis
  Software and Systems XVI. p.~127

\bibitem[\protect\citeauthoryear{Mevius et~al.,}{Mevius
  et~al.}{2016}]{Mevius_2016}
Mevius M.,  et~al., 2016, \mndoi [Radio Science]
  {https://doi.org/10.1002/2016RS006028}, 51, 927

\bibitem[\protect\citeauthoryear{{Michilli} et~al.,}{{Michilli}
  et~al.}{2018}]{Michilli2018Natur.553..182M}
{Michilli} D.,  et~al., 2018, \mndoi [\nat] {10.1038/nature25149}, \href
  {https://ui.adsabs.harvard.edu/abs/2018Natur.553..182M} {553, 182}

\bibitem[\protect\citeauthoryear{{Planck Collaboration} et~al.,}{{Planck
  Collaboration} et~al.}{2020}]{Plank2020A&A...641A...6P}
{Planck Collaboration} et~al., 2020, \mndoi [\aap]
  {10.1051/0004-6361/201833910}, \href
  {https://ui.adsabs.harvard.edu/abs/2020A&A...641A...6P} {641, A6}

\bibitem[\protect\citeauthoryear{Prochaska et~al.,}{Prochaska
  et~al.}{2019}]{Prochaska231}
Prochaska J.~X.,  et~al., 2019, \mndoi [Science] {10.1126/science.aay0073},
  366, 231

\bibitem[\protect\citeauthoryear{{Ravi} et~al.,}{{Ravi}
  et~al.}{2019}]{Ravi_2019}
{Ravi} V.,  et~al., 2019, \mndoi [\nat] {10.1038/s41586-019-1389-7}, \href
  {https://ui.adsabs.harvard.edu/abs/2019Natur.572..352R} {572, 352}

\bibitem[\protect\citeauthoryear{{Sault}, {Teuben}  \& {Wright}}{{Sault}
  et~al.}{1995}]{Miriad1995ASPC...77..433S}
{Sault} R.~J.,  {Teuben} P.~J.,   {Wright} M.~C.~H.,  1995, in {Shaw} R.~A.,
  {Payne} H.~E.,   {Hayes} J.~J.~E.,  eds,  Astronomical Society of the Pacific
  Conference Series Vol. 77, Astronomical Data Analysis Software and Systems
  IV. p.~433 (\mn@eprint {arXiv} {astro-ph/0612759})

\bibitem[\protect\citeauthoryear{Taylor, Carilli, Perley, of~the Pacific  \&
  (U.S.)}{Taylor et~al.}{1999}]{WhiteBooktaylor1999synthesis}
Taylor G.,  Carilli C.,  Perley R.,  of~the Pacific A.~S.,   (U.S.) N. R.
  A.~O.,  1999, Synthesis Imaging in Radio Astronomy II: A Collection of
  Lectures from the Sixth NRAO/NMIMT Synthesis Imaging Summer School Held at
  Socorro, New Mexico, USA, 17-23 June, 1998.
Astronomical Society of the Pacific conference series, Astronomical Society of
  the Pacific, \url {https://books.google.com.au/books?id=e23vAAAAMAAJ}

\bibitem[\protect\citeauthoryear{{Tendulkar} et~al.,}{{Tendulkar}
  et~al.}{2021}]{Tendulkar2021ApJ...908L..12T}
{Tendulkar} S.~P.,  et~al., 2021, \mndoi [\apjl] {10.3847/2041-8213/abdb38},
  \href {https://ui.adsabs.harvard.edu/abs/2021ApJ...908L..12T} {908, L12}

\bibitem[\protect\citeauthoryear{Walters, Weltman, Gaensler, Ma  \&
  Witzemann}{Walters et~al.}{2018}]{Walters_2018}
Walters A.,  Weltman A.,  Gaensler B.~M.,  Ma Y.-Z.,   Witzemann A.,  2018,
  \mndoi [ApJ] {10.3847/1538-4357/aaaf6b}, 856, 65

\bibitem[\protect\citeauthoryear{{Wright}}{{Wright}}{2006}]{Wright2006PASP..118.1711W}
{Wright} E.~L.,  2006, \mndoi [\pasp] {10.1086/510102}, \href
  {https://ui.adsabs.harvard.edu/abs/2006PASP..118.1711W} {118, 1711}

\makeatother
\end{thebibliography}

\end{document}